\documentclass[twoside,twocolumn,9pt]{article}
\usepackage{extsizes}
\usepackage[super,sort&compress,comma]{natbib} 
\usepackage[left=1.5cm, right=1.5cm, top=1.785cm, bottom=2.0cm]{geometry}
\usepackage{balance}

\usepackage{sectsty}
\allsectionsfont{\scshape}

\usepackage{graphicx} 
\usepackage{lastpage}
\usepackage{float}
\usepackage{fancyhdr}
\usepackage{fnpos}
\usepackage[english]{babel}
\usepackage{array}
\usepackage[T1]{fontenc}
\usepackage[usenames,dvipsnames]{xcolor}
\usepackage{setspace}
\usepackage{hyperref}
\usepackage[squaren]{SIunits}
\usepackage{titlesec}
\usepackage{upgreek}
\usepackage{xspace}
\usepackage{placeins}
\usepackage{amsmath,amssymb}

\newcommand{\Ro}{\ensuremath{R_\mathrm{0}}}
\newcommand{\tcoll}{\ensuremath{t_\mathrm{coll}}}

\newcommand{\hmax}{\ensuremath{h_\mathrm{max}}}
\newcommand{\phia}{\ensuremath{\phi_\mathrm{area}}}
\newcommand{\phic}{\ensuremath{\phi_\mathrm{c}}}
\newcommand{\We}{\ensuremath{{\rm We}}}
\newcommand{\Reynolds}{\ensuremath{{\rm Re}}}
\newcommand{\thetaR}{\ensuremath{{\theta_{\rm R}}}}
\newcommand{\thetaf}{\ensuremath{{\theta_{\rm f}}}}
\newcommand{\Psii}{\ensuremath{{\Psi_{\rm 1}}}}

\begin{document}

\twocolumn[
  \begin{@twocolumnfalse}
\vspace{3cm}

\begin{center}

    \noindent\Huge{\textbf{\textsc{Impact and spreading dynamics of a drop of fiber suspension on a solid substrate}}} \\
    \vspace{1cm}

    \noindent\large{Sreeram Rajesh,\textit{$^{a}$} Alban Sauret\textit{$^{b,c}$}} \\

    \vspace{5mm}


    \vspace{1cm}
    \textbf{\textsc{Abstract}}
    \vspace{2mm}

\end{center}

\noindent\normalsize{

\textit{\it Hypothesis:} The presence of non-Brownian spherical particles dispersed in a liquid modifies the impact and spreading dynamics of a drop on a hydrophilic substrate. The difference in spreading dynamics is attributed to the increase in the viscosity of the suspension caused by the presence of the particles. Similarly, the presence of anisotropic non-Brownian particles, such as fibers, also increases the bulk viscosity of the suspension. In addition to the diameter \(D\) of the fiber, the length \(L\), which determines the aspect ratio \(A = L/D\), is crucial in controlling the viscosity of a fiber suspension. Therefore, we hypothesize that the drop impact of fiber suspensions with different volume fractions will result in a similar modification of the spreading dynamics.

\smallskip

\noindent \textit{Experiment:} To investigate the impact and spreading dynamics, we prepare suspensions of fibers with an aspect ratio \(A = 12\) at different volume fractions. These volume fractions span the dilute, semi-dilute, and dense concentration regimes. Additionally, we conduct a subset of experiments with aspect ratios \(A = 4\) and \(A = 20\). Furthermore, we characterize the thickness of the resulting droplet film, as well as the coating and orientation of fibers after the spreading dynamics reach a steady state.

\smallskip

\noindent \textit{Findings:} The presence of fibers significantly influences the spreading dynamics and final size of the droplet on the hydrophilic substrate. Notably, the resulting droplet size after spreading decreases as the volume fraction of fibers in the suspension increases. To rationalize these results, we use a modified equation, originally developed for spherical particles, which incorporates the viscosity of the suspension. Additionally, we observe an increase in the splashing of the droplet during spreading when increasing the Weber number and the volume fraction. Furthermore, we show that as the volume fraction increases, the final thickness of the droplet increases, and the resulting fiber coating becomes less uniform. We also highlight the secondary influence of fiber geometry on the coatings, such as the overlap of fibers, which further affects the coating uniformity. Despite these geometry-induced modifications, the radial orientation of the fibers remains isotropic across all volume fractions considered in this study.
} \\

 \end{@twocolumnfalse} \vspace{0.6cm}
 
 ]

\makeatletter
\renewcommand{\thefootnote}{}
\renewcommand*{\@makefnmark}{}
\footnotetext{\textit{$^{a}$~Department of Mechanical Engineering, University of California, Santa Barbara, California 93106, USA; E-mail: sreeram@ucsb.edu }}
\footnotetext{\textit{$^{b}$~Department of Mechanical Engineering, University of Maryland, College Park, Maryland 20742, USA; E-mail: asauret@umd.edu}}
\footnotetext{\textit{$^{c}$~Department of Chemical Engineering, University of Maryland, College Park, Maryland 20742, USA}}
\makeatother


\section{Introduction}
The impact, spreading, and final deposition of liquid droplets on solid substrates are encountered in a wide range of natural and industrial situations, from raindrops falling on soil to suspension drops in inkjet printers. Various aspects of drop impact on solid surfaces have been studied, such as the early dynamics after contact, the role of substrate wettability, the maximum spreading diameter, the splashing threshold, or the force exerted on the substrate \cite[see, \textit{e.g.},][]{josserand2016drop,cheng2022drop}. Such studies have greatly improved our understanding of the impact of droplets of Newtonian fluids. However, many industrial processes involve capillary flows of complex fluids such as polymer solutions \cite[\textit{e.g.},][]{fardin2022spreading, rajesh2022transition, rajesh2022pinch}, particulate suspensions \cite[\textit{e.g.},][]{furbank2004experimental, thievenaz2022onset}, and colloidal particles \cite[][]{bertola2015impact}. The numerous applications of drop impacts have motivated many recent studies with more complex rheology \cite{shah2024drop}. For instance, on the impact of shear thickening and shear thinning fluids on a hydrophobic surface \cite{guemas2012drop}, numerical simulations of the impact of viscoelastic liquid droplets \cite[][]{wang2017impact}, the impact of yield-stress fluids on a flat surface \cite[]{luu2009drop}, and drop impact on thin films \cite[][]{sen2020viscoplastic}. Such studies have extended our understanding of drop impact to homogeneous complex fluid systems.

\smallskip

However, studies investigating the effect of discrete particles during the impact and deposition of suspension drops on substrates remain scarce. This topic is important because the deposition of droplets of particulate suspensions occurs in processes such as inkjet printing \cite[][]{tekin2008inkjet, lohse2022fundamental}, 3D printing of fiber composite matrices \cite[][]{parandoush2017review}, and drug delivery to designated parts of the respiratory system \cite[][]{lopez2019printer}. The deposition of fiber particulates such as asbestos is also of environmental concern, as the pollutants can be deposited in the lungs \cite[see][]{asgharian1989deposition}. Past studies have considered the role of non-Brownian particles dispersed in a fluid during capillary flows. For instance, particles have been shown to modify the pinch-off of droplets \cite{furbank2004experimental,bonnoit2012accelerated,chateau2018pinch,thievenaz2021pinch,thievenaz2021droplet} or thin-film coating processes \cite{gans2019dip,jeong2022dip,jeong2022particulate}. The role of spherical, non-Brownian particles has also been investigated for the drop impact on flat surfaces \cite[][]{nicolas2005spreading, grishaev2015complex, grishaev2017impact}, and on cylindrical targets \cite[][]{raux2020spreading}. The studies on drop impacts of particle-laden fluids have characterized the morphology of the suspension droplet upon impact, the splashing threshold, and the final particle distribution. However, the influence of anisotropic non-Brownian particles, such as fibers, during the impact, spreading, and coating of a suspension drop on a solid substrate has received much less attention and remains unclear. In the present study, we experimentally investigate the impact of a drop of fiber suspension on a hydrophilic glass surface to address these questions. 

\smallskip

Interfacial flows of non-Brownian fiber suspensions exhibit complex multiphase dynamics arising from interactions between discrete fibers and the surrounding liquid-air interface \cite{solomon2010microstructural, chateau2018pinch}. In many industrial applications, capillary interactions of fibers play a crucial role \cite{fuller2012complex, jeong2023deposition}. When fibers are introduced at capillary scales, \textcolor{black}{comparable to the characteristic length scale of the flow,} their random packing within the liquid is modified, influencing the overall suspension dynamics. This is particularly pronounced at larger volume fractions, potentially inducing fiber ordering or resulting in fibers protruding through the liquid-air interface. In applications involving additive manufacturing and spray-coating, capillary interactions influence both the coated area and fiber orientation, which are crucial to material strength \cite{truby2016printing, turton2005scale}. \textcolor{black}{In bioprinting of organs, capillary-scale interactions of deformable anisotropic particles are relevant to cell extrusion, where precise cell alignment is essential \cite{murphy20143d}. Similarly, capillary-driven flows are also important in applications such as self-healing gels, localized drug release and delivery, fibrous porous water treatment membranes, and industrial-grade separations, where extrusion-based methods are used to deposit fibers and anisotropic particles \cite{hia2016electrosprayed, haase2017multifunctional, wang2021wireless}.} Another well-studied example is capillarity-driven flows of particles toward the contact line during the evaporation of colloidal suspensions, affecting uniformity and coverage \cite{hu2002evaporation, Deegan1997, Dufresne2003}. The present study aims to identify key quantities relevant to industrial applications involving the deposition of droplets of fiber suspensions and examine how parameters such as volume fraction and fiber geometry influence these quantities.

\smallskip

While the spreading dynamics, final droplet size, shape, and splashing threshold have been studied for spherical particles, the effects of fibers remain largely unexplored \cite[][]{nicolas2005spreading, pasandideh1996capillary, grishaev2015complex, grishaev2017impact}. The approach used in this study successfully quantifies these parameters for fiber suspensions. Notably, we introduce the concept of area fraction to quantify the final surface area coated by the fibers following the droplet impact, enabling a description of the effects introduced by the presence of anisotropic particles within the droplet. Additionally, we develop an in-house routine to quantify the radial orientation of fibers in the deposited droplet, with the capability to distinguish overlapping fibers accurately. Our investigations provide a framework for quantifying key factors associated with capillary flows and interfacial effects in the droplet deposition of non-Brownian fiber suspensions. Importantly, we address a critical gap in the literature on droplet deposition involving anisotropic particles.

\smallskip

The paper is organized as follows: in \S \ref{sec:exp_methods}, we present the experimental techniques used in this study. In \S \ref{sec:Spreading_d}, we discuss the dynamics of the droplet as it impacts and spreads on a surface, the resulting thickness of the deposited film, and map the splashing after the drop impact. Following this, we present a discussion on the coating of the deposited fibers and the distribution of their orientations in \S \ref{sec:coating_orientation}. We then draw concluding remarks in \S \ref{sec:conclusions}.


\begin{figure}
  \centerline{\includegraphics[width=0.49\textwidth]{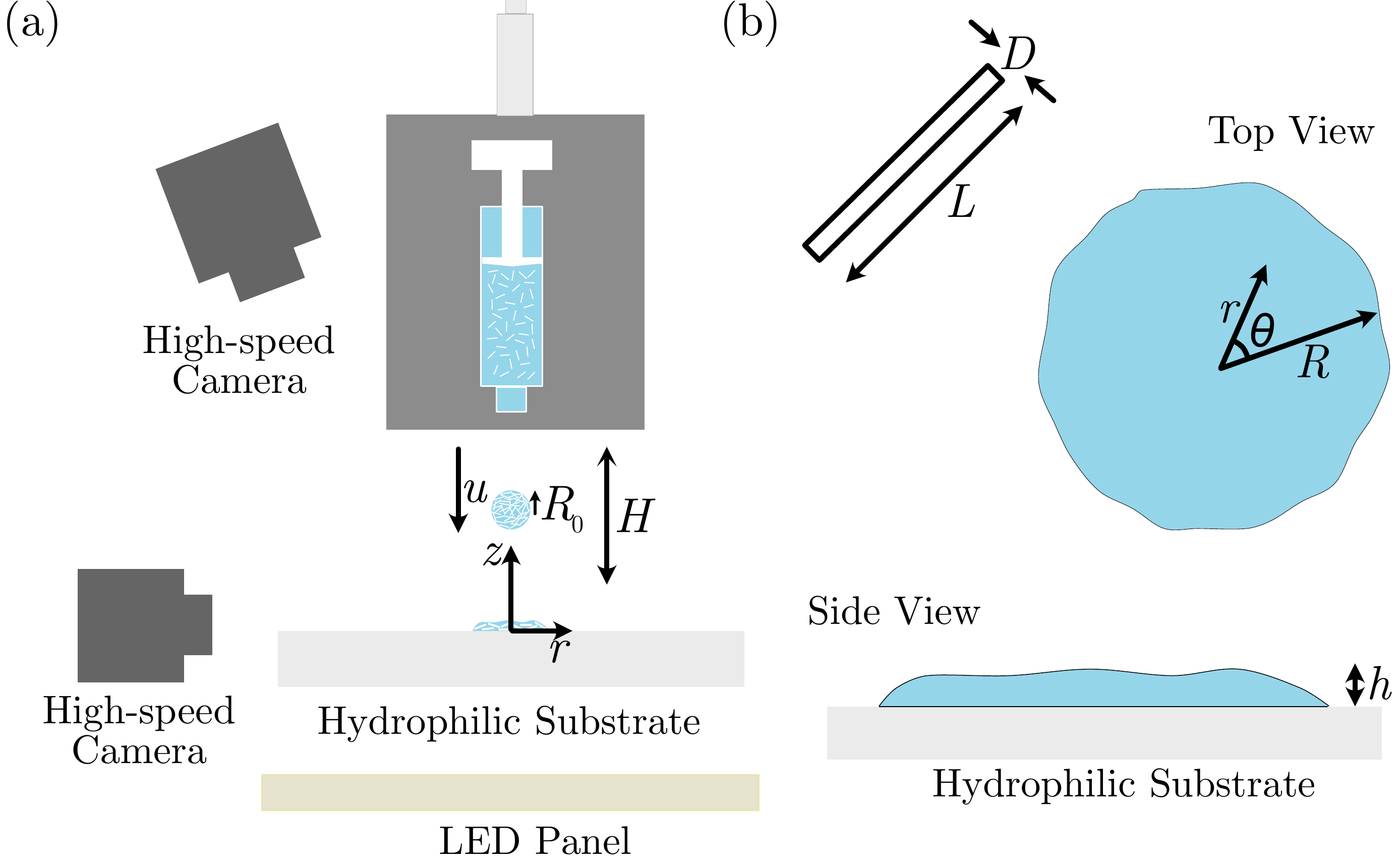}}
  \caption{(a) Schematic of the experimental setup. (b) Top and side views of the geometry of the droplet as it spreads. \textcolor{black}{The fibers have cylindrical cross-section of diameter $D$ and a length $L$.}}
\label{fig:drop_impact_schema}
\end{figure}

\begin{figure*}
  \centerline{\includegraphics[width=0.9\textwidth]{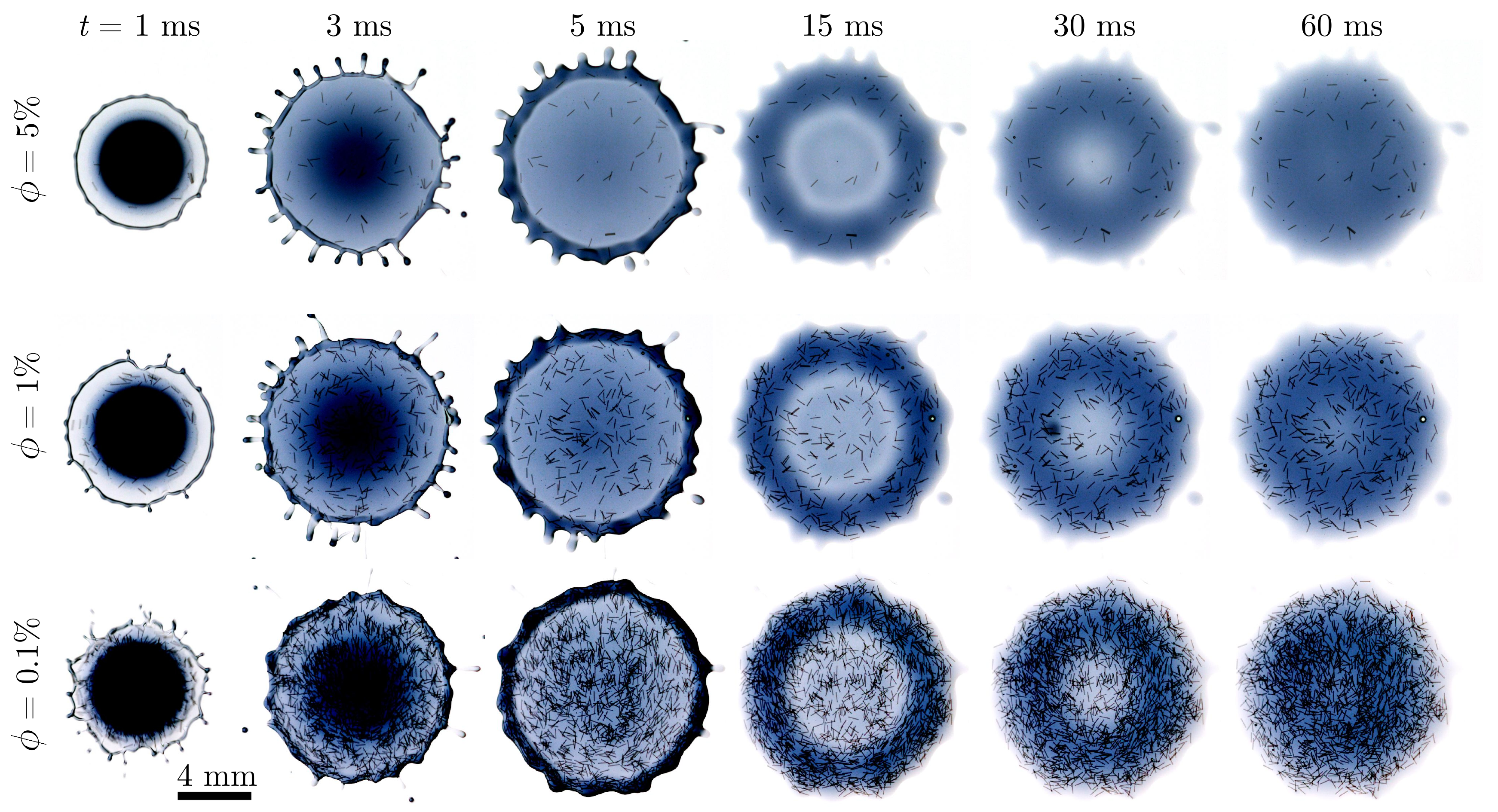}}
  \caption{Snapshots of time-evolution of the suspension droplet upon impact for volume fraction $\phi = 0.1\%$, $\phi = 1\%$, and $\phi = 5\%$ for fibers of length $L = 600\,\rm \mu m$ and diameter $D = 50\,\rm \mu m$.}
\label{fig:drop_impact}
\end{figure*}

\section{Experimental Methods}
\label{sec:exp_methods}

The experimental setup, shown in figure \ref{fig:drop_impact_schema}(a), consists of a syringe filled with the fiber suspension mounted vertically on a micrometer translation stage. The quasi-static extrusion of the fiber suspension from a nozzle of diameter $3 \rm \, mm$ leads to a pendant drop. The drop detaches when gravity overcomes the capillary forces, leading to droplets of diameter $d_{\rm 0} = 2R_{\rm 0} =4 \pm 0.1 \, {\rm mm}$. The droplet is released from a height $ H = $ 10, 14, and 18 ${\rm cm}$, resulting in impact velocities $U =$ 1.4, 1.65, and 1.9 ${\rm m/s}$, and thus a timescale of impact $t_{\rm i} =d_{\rm 0}/U \approx 2$ - $ 4 \rm \, ms$. More details on the free-fall dynamics of the droplet are provided in \S S1 of the supplementary material. To ensure consistent results, we performed seven realizations of the experiment for each parameter investigated. The substrate on which the droplet impacts is a hydrophilic glass surface. We prepare the substrate by polishing it with an abrasive compound (3M, Cerium Oxide), ensuring a consistent surface for all experiments and a small contact angle, smaller than $5^{\rm o}$ for the solvent considered.

\smallskip

The suspension is made of non-Brownian nylon fibers dispersed in a density-matched solvent made of a mixture of water and glycerol (40\%/60\% by weight) of density $\rho = 1152 \, {\rm kg/m^3}$ (measured with Anton-Paar DMA 35), dynamic viscosity $\eta_{\rm f} = 9.5 \, {\rm mPa.s}$ (measured with Anton Paar MCR 302), and surface tension $\gamma = 67 \, {\rm mN/m}$ (measured with Attension Tensiometer). The solvent has a contact angle $\theta_{\rm c} < 5^{\rm o}$ on the glass substrate. Note that the presence of particles does not modify the density or the surface tension of the suspension (\cite[See Ref.][]{pelosse2023probing, couturier2011suspensions}). We disperse the nylon fibers of various aspect ratios $A = L/D$ at different volume fractions $\phi$ in the solvent. Most of the experiments are performed with fibers of diameter $D \approx 50 \rm \, \mu m$ and length $L \approx 600 \,\mu {\rm m}$, resulting in $A = 12$. We also present some results for a smaller set of volume fractions for fibers of aspect ratios $A = 4$ and $A =20$ and the same diameter $D = 50 \rm \, \mu m$ (See figure S2 in supplementary materials for the size distribution of the fiber lengths). Overall, we use a volume fraction $\phi \in [0.1,\,20]\%$, depending on the aspect ratio. Indeed, for too large aspect ratios or volume fraction, the fibers may clog the nozzle \cite[][]{croom2021mechanics,dincau2023clogging}. For the range of volume fractions used here, we ensured that the volume fraction of fibers in the extruded drop is similar to the volume fraction in the nozzle, \textit{i.e.}, there is no self-filtration at the syringe nozzle \cite[][]{haw2004jamming} (see supplementary material, figure S3). 

\smallskip

In the present work, we considered moderate Weber numbers $\We = \rho U^2 d/\gamma = $ 135, 190, and 240. The Reynolds number of the solvent droplet upon impact is defined as $\Reynolds_{\rm f} = \rho U d/\eta_{\rm f}$ and takes the values 680, 800, and 920. Alternatively, we could define a suspension Reynolds number as $\Reynolds(\phi) = \rho U d/\eta(\phi) = \Reynolds_{\rm f}\,\eta_{\rm f}/\eta(\phi)$, where $\eta(\phi)$ is the viscosity of the fiber suspension. The impact and spreading dynamics are recorded from the top and the side at $10,000$ frames per second (VEO Phantom 640), as shown in figure \ref{fig:drop_impact}.

\smallskip

For an aspect ratio $A = 12$, the limits of the dilute, semi-dilute, and dense regimes are $\phi \ll 0.54\%$, $ 0.54\% \leq \phi \ll 6.5\%$, and $\phi \geq 6.5\%$, respectively, based on the limits provided by Butler \textit{et al.} \cite[][]{butler2018microstructural} (Further details provided in \S S1, supplementary material). In the semi-dilute or dense regime, the hydrodynamic and fiber-fiber interactions cannot be neglected \cite[][]{shaqfeh1990hydrodynamic}. Different empirical correlations have been proposed to describe the viscosity of fiber suspensions $\eta(\phi)$. Here, in the range of volume fraction considered, we use the Maron-Pierce correlation, giving the steady-state shear viscosity of fiber suspension \cite[][]{bounoua2019shear}:
\begin{equation}
\label{eqn:maron_pierce}
  \eta = \eta_{\rm 0}(1-\phi/\phi_{\rm c})^{-2},
\end{equation}
where $\phi_{\rm c}$ is the critical volume fraction at which the viscosity diverges, and depends on the fiber aspect ratio. We should emphasize that other empirical correlations would have been possible for the present fiber suspensions. \textcolor{black}{For instance, Tapia \textit{et al.}\cite{tapia2017rheology} proposes $\eta = \eta_{\rm 0}(1-\phi/\phi_{\rm c})^{-0.9}$ for fibers of volume fraction $\phi > 25\%$. However, for the range of volume fraction considered here ($\phi \leq 10\%$), we use the Maron-Pierce correlation, similar to the range of $\phi$ explored by Bounoua \textit{et al.}\cite[][]{bounoua2019shear}. Further, the Maron-Pierce model has been previously found to work well to describe the dip-coating of substrates with fibers \cite{jeong2023deposition}}.


\begin{figure*}
  \centerline{\includegraphics[width=\textwidth]{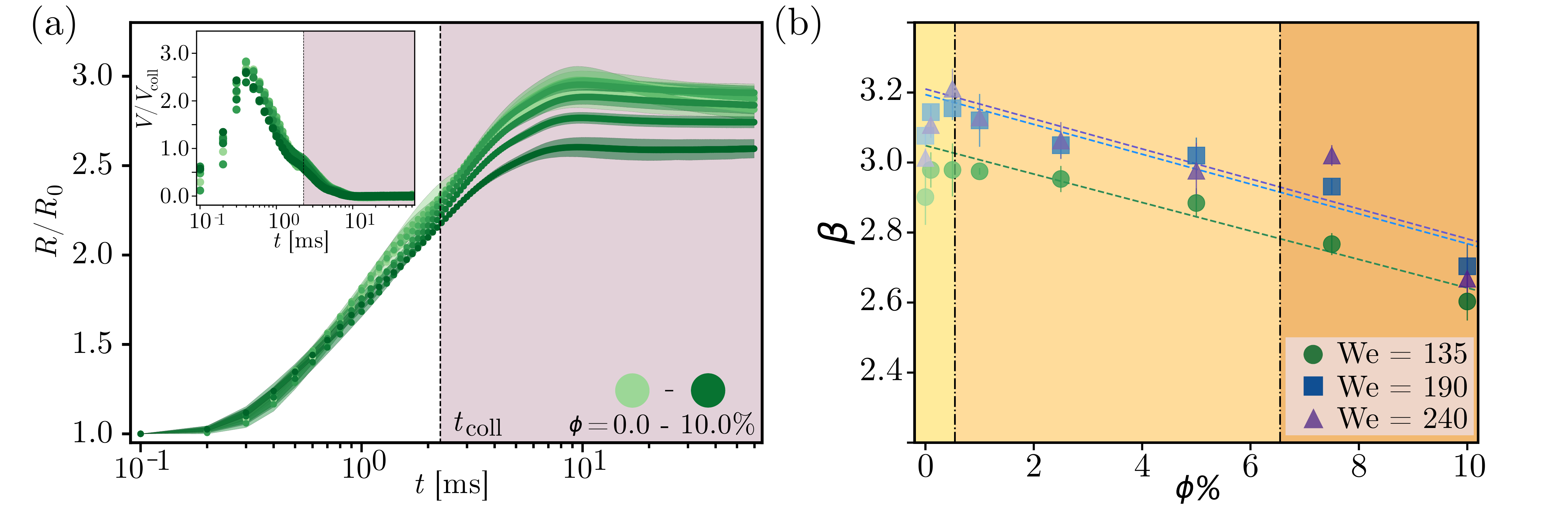}}
  \caption{(a) Time evolution of the rescaled radius $R/\Ro$ for varying volume fraction $\phi \in \left[0, 10\right]\%$ of fibers ($A=12$) and $\We = 150$. The shaded region shows the later stage of spreading once the droplet has fully impacted the surface ($t$ > \tcoll). Inset: time-evolution of the rescaled spreading velocity $V/V_{\rm coll}$. (b) Spreading factor $\beta = {\rm max}(R/\Ro)$ for suspensions of  $A = 12$ fibers with $\phi \in \left[ 0, 10 \right]\%$ and $\We = 135-240$. \textcolor{black}{The dashed lines indicate fits using equation \ref{eqn:beta_fit_linear}.} The colored regions indicate the dilute, semi-dilute, and dense regimes (from left to right)}
\label{fig:Spreading_d}
\end{figure*}

\section{Results and discussion}
\subsection{Spreading dynamics of fiber suspensions}
\label{sec:Spreading_d}

\subsubsection{Initial spreading of the suspension droplet}
\label{subsec:radial_spread}

\smallskip

The droplet is released at a height $H$ and impacts the substrate with a velocity $U$. Figure \ref{fig:Spreading_d}(a) shows the time-evolution of the rescaled radius $R(t)/R_{\rm 0}$ of the outer edge of the droplet for $A = 12$ and varying the volume fractions $\phi \in [0,\,10]\%$ at $\We = 135$. We consider here the spreading dynamics of the droplet after the initial contact with the surface at $t = 0 \rm \, ms$ and until the radial spread reaches a steady-state value around $t \simeq 100 \rm \, ms$. While the droplet continues to relax beyond this time, \textcolor{black}{the increase in the droplet size is about $25\%$ from the maximum droplet size over $ t \simeq 30 \, {\rm s} \gg 100 \, {\rm ms}$. The long-term evolution of the drop is driven by capillarity and gravity, unlike the early-time inertia-driven spreading dynamics. Additionally, the evaporation of the solvent is also relevant for the long-term droplet size evolution ($t \geq 30 \, \rm s$).} Therefore, a detailed description of the long-term evolution of the drop is beyond the scope of the present study.

\smallskip

During the initial spreading, \textit{i.e.}, for $t<\tcoll$, the radius $R(t)$ rapidly increases, as seen in the unshaded region in figure \ref{fig:Spreading_d}(a). Here, $\tcoll$ is the collision time defined as $\tcoll = 2\Ro/U$ \cite[][]{pasandideh1996capillary, eggers2010drop}. The value of $\tcoll \approx 2.5 \rm \, ms$ for $\We = 135$ is shown by the dotted lines in figure \ref{fig:Spreading_d}(a). This timescale agrees well with our experimental observation of when the spreading radius $R$ approaches a steady value. The evolution of the dimensionless radius $R(t)/\Ro$ shows that the addition of fibers slightly slows down the initial spreading. Indeed, as we increase $\phi$, the macroscopic viscosity $\eta(\phi)$ increases in the presence of particles, resulting in larger viscous dissipation.

\smallskip
The spreading velocity, non-dimensionally defined as $V/V_{\rm coll}$, where $V={\rm d}R /{\rm d}t$ and $V_{\rm coll}=R_0/\tcoll$, is reported in the inset of figure \ref{fig:Spreading_d}(a) for the same experimental parameters as in the main panel. Following the impact, the spreading velocity reaches a peak before the spreading slows down due to viscous effects. The viscous dissipation arises from two contributions: wall friction on the substrate and the presence of particles. As expected, the peak velocity decreases slightly when increasing $\phi$, due to the increase in macroscopic viscosity.
\medskip

\subsubsection{Final spreading radius}
\label{subsec:final_spread}

After the collision ($t > \tcoll$), the droplet spreads to a maximum radius, as seen in figure \ref{fig:Spreading_d}(a), before slightly relaxing to a final shape for $t \gg \tcoll$. As the droplet spreads, the kinetic energy from the impact is dissipated through viscous effects, and $R/\Ro$ plateaus toward a constant value. To quantify the influence of $\phi$ on the maximum spreading radius, we define the spreading factor $\beta = {\rm max}(R/\Ro)$  \cite{nicolas2005spreading}. We report in figure \ref{fig:Spreading_d}(b) the evolution of $\beta$ measured from our experiments for the $A = 12$ fibers at $\We = 135$, $190$, and $240$ and increasing volume fraction. The shaded regions from light to dark represent dilute, semi-dilute, and dense regimes \cite[][]{butler2018microstructural}. In the dilute regime $ 0\% < \phi < 1\%$, we note that $\beta$ slightly increases as volume fraction increases. However, in the semi-dilute and dense regimes, $\phi \gtrsim  1\%$, the final radius decreases as $\phi$ increases, as expected from the increase in viscous effects with $\phi$. We also observe similar results for the $A = 4$ and $A = 20$ fibers at different $\We$ (figure S4 in the supplementary material).

\smallskip

We can try to rationalize the evolution of $\beta(\phi)$ using a model developed by Pasandideh \textit{et al.} \cite[][]{pasandideh1996capillary} for particle-free Newtonian liquids, and extended by Nicolas \cite[][]{nicolas2005spreading} for suspensions of neutrally buoyant spherical particles. When an initial suspension drop of volume $\Omega = 4\,\pi {\Ro}^3/3$ impacts a solid substrate from a height $H$, the total energy of the drop just before the impact is purely due to the kinetic energy gained from the free fall, $E_{\rm K} = 4\,\pi \rho {\Ro}^3 U^2/3$. After the impact, the spreading droplet dissipates energy through viscous losses, $W_{\rm d} = 4\,\alpha\,\pi\,\eta\, {\Ro}^2\beta^4$, where $\alpha$ is a fitting parameter of order 1. \textcolor{black}{Similar viscous dissipation controlling bounding and merging behavior has also been quantified for liquid droplets impacting on thin liquid films \cite{tang2018bouncing}.} For the solvent, an energy balance provides the expression for the spreading factor \cite[][]{pasandideh1996capillary}: 
\begin{equation}
    \beta = \left(\frac{\Reynolds_{\rm f}}{12\alpha}\right)^{1/4}
\end{equation}

For suspensions of particles, the Reynolds number ${\Reynolds}(\phi)$ is related to the Reynolds number ${\Reynolds}_{\rm f}$ of the solvent through $\Reynolds(\phi) = {\rm Re}_{\rm f}/\eta_{\rm r}(\phi)$. As defined in equation \ref{eqn:maron_pierce}, the relative viscosity is given by $\eta_{\rm r} = \eta(\phi)/\eta_{\rm 0} =(1-\phi/\phi_{\rm c})^{-2}$, such that the expression for the spreading factor for the droplet of fiber suspension becomes:
\begin{equation}
\label{eqn:beta_fit}
    \beta = \left(\frac{\Reynolds_{\rm f}}{12\alpha}\right)^{1/4}\left(1-\frac{\phi}{\phi_{\rm c}}\right)^{1/2}
\end{equation}

\textcolor{black}{Furthermore, when $\phi \ll \phic$ we can approximate equation \ref{eqn:beta_fit} as a linear expression:}
\begin{equation}
\label{eqn:beta_fit_linear}
    \beta = \left(\frac{\Reynolds_{\rm f}}{12\alpha}\right)^{1/4}\left(1 - \frac{1}{2}\frac{\phi}{\phi_{\rm c}}\right)
\end{equation}

We use equation \ref{eqn:beta_fit_linear} to fit $\beta$ measured from the experiments, as shown in figure \ref{fig:Spreading_d}(b), which captures $\beta(\phi)$ well in the semi-dilute and dense regime (\textit{i.e.}, $\phi > 1\%$) for $A = 12$ fibers at different $\We$. For fibers, the maximum random packing fraction $\phic$ depends on aspect ratio $A$ (\cite[See Ref.][]{bounoua2019shear}). For $A = 12$, we consider \phic $ = 37.5\%$, which provides \textcolor{black}{the viscosity of the suspension in the range $\eta \in [9.5, \, 17.66] {\rm \, mPa.s}$ for $\phi \in [0, \, 10]\%$, and} a fitting parameter $\alpha \approx 0.76 \pm 0.03$ that depends weakly on $\We$. Similar fits for the $A = 4$ and $A = 20$ fiber suspensions are provided in \S S4 of the supplementary material. The values of $\phic$ used here are based on estimates from the work of Bounoua \textit{et al.} \cite[][]{bounoua2019shear} and describe well the evolution of $\beta$ measured in this study. Nevertheless, $\phi_{\rm c}$ remains an estimate, and detailed measurement of the rheology of fiber suspensions at different aspect ratios would allow us to obtain $\phic(A)$.

\smallskip

The values of $\alpha$ estimated from our experiments are close to the values reported in Ref. \cite{nicolas2005spreading} ($\alpha \approx 0.89$ for spherical particles on a hydrophilic solid substrate). The larger values of $\alpha$ observed for $A = 4$ and $A = 20$ noted in \S S4 are likely due to a smaller sample set of $\phi$ (only three values of $\phi$ were considered, as shown in figure S4). However, this confirms our hypothesis on the role of viscous dissipation $\eta(\phi)$ on the final size of the suspension droplet, despite the differences in aspect ratio $A$. In summary, at first order, the spreading behavior of the fiber suspension droplets on a hydrophilic substrate can be captured through the modification of the effective macroscopic viscosity, and one can neglect the local heterogeneity brought by the particles. This observation is reminiscent of the similar configuration with spherical particles. Therefore, the main effect of the geometry of the fibers here is to change the value of $\phic$ and, thus, the viscosity at a given volume fraction, and one can neglect the local heterogeneities brought by the fibers during this phase.

\medskip

\subsection{Splashing criterion for fiber suspensions}

\begin{figure}[!t]
    \centerline{\includegraphics[width=0.4\textwidth]{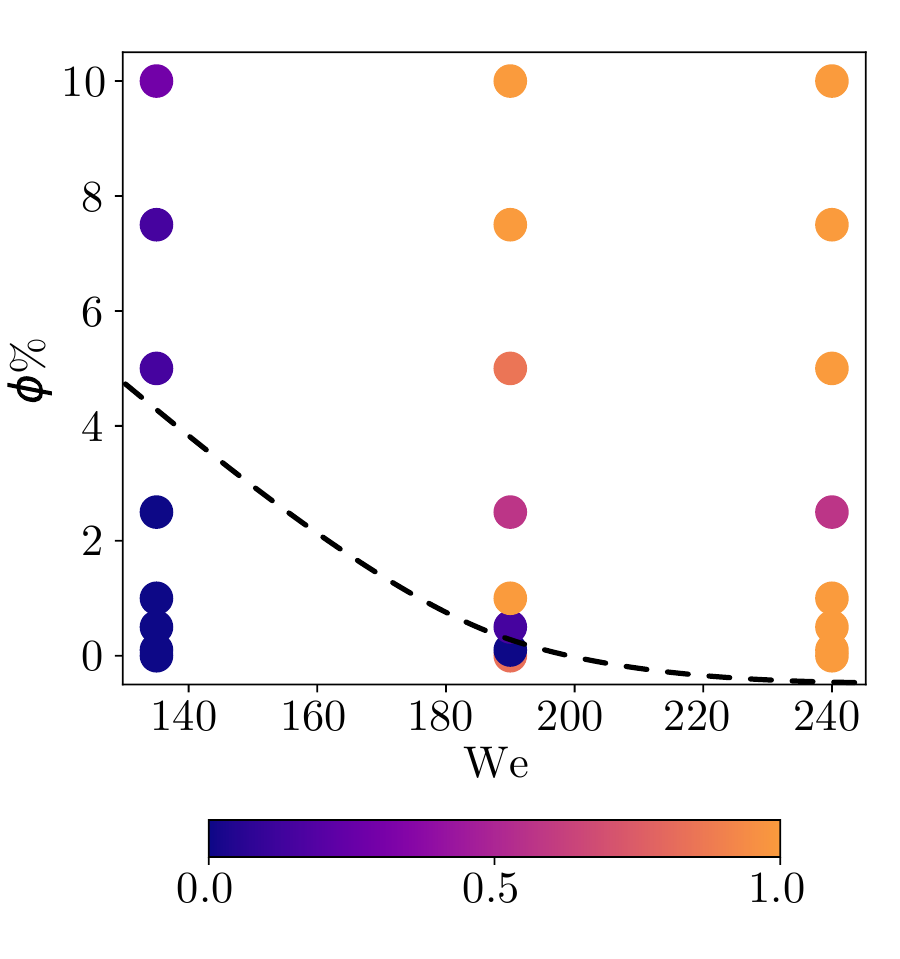}}
  \caption{Diagram showing the splash/no splash regions for the $A = 12$ fibers as a function of $\phi$ and $\We$. The color map corresponds to the normalized number of droplets ejected. The dashed lines are a guide for the eye, demarcating the splash/no splash boundary.}
\label{fig:splash_criterion}
\end{figure}

\smallskip

Understanding the splashing of suspension droplets is essential for good predictions of the final size and shape of the deposited droplet \cite{josserand2016drop}. Various splash criteria relating the droplet Reynolds number, $\Reynolds$, and Weber number, $\We$, have been proposed for a liquid impacting on a substrate \cite[]{moreira2010advances, marengo2011drop}. Modified versions of these expressions have also been investigated for suspensions of spherical particles \cite[]{yarin2006drop, mundo1995droplet, cossali1997impact,grishaev2015complex}. However, such models fail to capture the splash/no splash observations in the present study. A modified criterion for splashing in the drop impact of suspensions has been proposed by defining a particle-based Weber number \cite{peters2013splashing}. However, this model does not address the role of $\phi$. 

\smallskip

In figure \ref{fig:splash_criterion}, we show the regime map for the splashing of $A = 12$ fibers with $\phi \in [0, 10\%]$. The color map corresponds to the normalized number of droplets ejected. The dashed lines are a guide for the eye, demarcating the splash/no splash boundary. As observed, the criterion not only depends on $\We$ and $\Reynolds$, but also on $\phi$. At $\We = 135$, we observe splashing only when $\phi > 2.5\%$. For $\phi > 2.5\%$, as $\phi$ increases, the number of droplets ejected also increases. The threshold decreases with $\We$. For $\We = 190$, we see splashing when $\phi = 0\%$ and for $\phi > 0.1\%$. At $\We = 240$, we observe splashing for all values of $\phi$. Similar splash/no splash observations have been previously reported for spherical particles \cite{grishaev2015complex}, where splashing was observed with an increase in both $\We$ and $\phi$. This work, however, did not describe the number of droplets ejected. We conclude from our observations that the splashing behavior of suspension droplets is influenced by multiple factors, including $\We$, $\Reynolds$, and $\phi$. For fiber suspensions, additional factors such as aspect ratio $A$ may also modify this behavior. From a limited number of tests, we observe an increase in splashing with $\We$ for $A = 4$ and $A = 20$ fibers. However, a larger data set is necessary to build an accurate model on the influence of $A$ on splashing and is beyond the scope of the study, where our primary focus is on the final deposited droplet. This is considered in the following sections.

\medskip


\begin{figure*}[!h]
  \centerline{\includegraphics[width=0.96\textwidth]{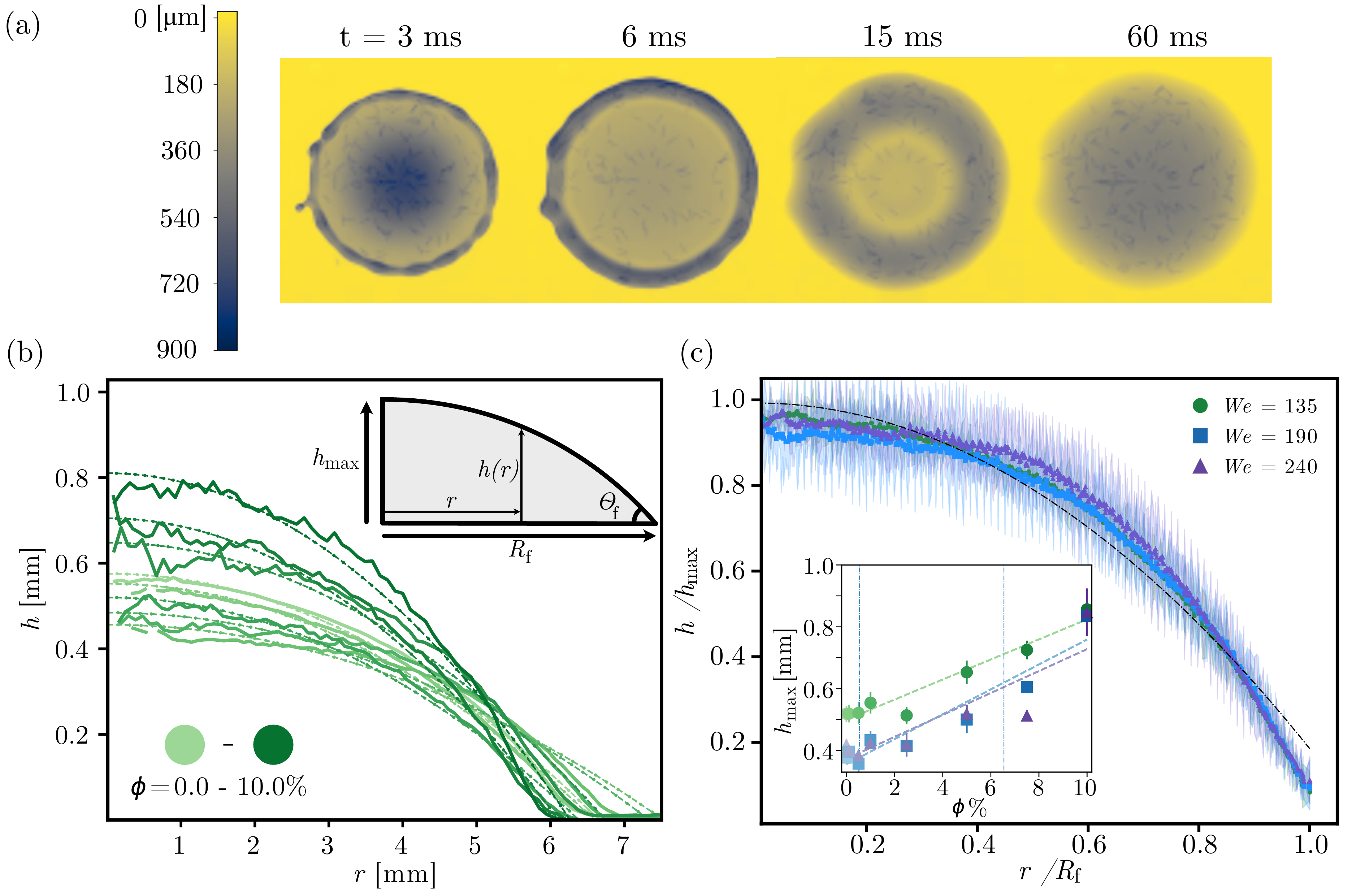}}
  \caption{(a) Thickness profiles of a fiber suspension droplet of volume fraction $\phi = 0.5\%$ ($A = 12$) evaluated using absorption spectroscopy as it spreads on the substrate. (b) Final droplet thickness profiles (Solid green lines, $t \simeq 100 \, {\rm ms}$) for suspensions of fibers of aspect ratio $A=12$ and volume fractions $\phi \in \left[0-10\right]\%$ (from light to dark green) and $\We = 135$. The dotted lines are fitted profiles, assuming the final profiles have the shape of a spherical cap (see Equation \ref{eqn:thickness}). Inset: Schematic of the steady-state droplet morphology and definition of the different physical quantities. (c) Rescaled mean thickness profiles for $A = 12$ fiber suspensions at different volume fractions. The solid lines show a smoothened signal overlaid on a shaded region with the raw mean thickness profile. Inset: Maximum thickness $\hmax(\phi)$ of the deposited droplet at different $\We$. The error bars are obtained from five realizations of each experiment. The dotted lines are given by Eq. (\ref{eqn:hmax}).} 
\label{fig:thickness}
\end{figure*}


\subsection{{Final thickness profile of the droplet}}
\label{subs:thickness}

Quantifying and predicting the thickness profile of droplets on a surface is important in coating applications, inkjet printing, and additive manufacturing \cite[][]{ lohse2022fundamental, truby2016printing, sauret2019capillary, siemenn2022machine}. Various methods can be used to measure or estimate the thickness of the liquid film, such as imaging from the side or estimating the average thickness from the droplet weight and radius. Such methods, however, have the limitation that the thickness of the droplet can only be estimated along a fixed plane or on an average. An accurate way to measure the thickness for the entire drop is through a light absorption technique, where the intensity transmitted by a dyed liquid film is mapped to a previously calibrated thickness \cite{vernay2015free}. This technique was initially considered for pure liquid \cite{vernay2015free} and later used with suspensions of spherical particles \cite{raux2020spreading}. Here, we adopt this technique for fiber suspensions.

\smallskip

Before we prepared the suspension, we added $1.2 \rm \, g\,L^{-1}$ of Nigrosin (Sigma-Aldrich), \textcolor{black}{a black-dye commonly used for biological samples,} to the \textcolor{black}{solvent}. \textcolor{black}{Nigrosin has been shown to not alter the surface tension or the viscosity of the liquid \cite[][]{wang2017drop}.} The details of the calibration and method are provided in \S S5 of the supplementary material. The calibration allows us to convert the intensity $I$ measured at any location and time by the high-speed camera to the local liquid thickness (from the calibration parameter and the reference intensity $I_{\rm 0}$). We show an example of the time evolution of the 3-dimensional thickness profile extracted from the procedure for a fiber suspension of volume fraction $\phi = 0.5\%$ and $A = 12$ in figure \ref{fig:thickness}(a).

\smallskip

From our observations on droplet spreading dynamics, shown in figure \ref{fig:Spreading_d}(a), the droplet reaches a steady state of spreading around $t \simeq 100 \rm \, ms$. Using the technique described previously, we extract the thickness profile of the droplet in this final state for $A = 12$ at different $\phi$ and ${\rm We}$. We show the azimuthally averaged profiles for one realization of the experiment in figure \ref{fig:thickness}(b) (solid green lines) for $0 \leq \phi \leq 10\%$ and $\We = 135$. The shape of the droplets in this state can be described, at first order, as a spherical cap. This is a reasonable assumption since the maximum height of the droplets on the substrate is of order $h \sim 1 {\, \rm mm}$, smaller than the capillary length $\ell_{\rm c} = 2.3\,{\rm mm}$ and the Weber number is small. Hence, capillarity dominates. Note that such an assumption is expected to work well in the limit of vanishing Weber numbers and deviate at larger Weber numbers. The assumption of a spherical cap has been previously used to study the evaporation of a sessile droplet deposited on a hydrophilic surface \cite[]{hu2002evaporation}. A schematic of the drop as a spherical cap is shown in the inset of figure \ref{fig:thickness}(b). The equation for a spherical cap is:
\begin{equation}
    h(r) = \left[ \left(\frac{R_{\rm f}}{\sin{\thetaf}} \right)^{2} - r^2\right]^{1/2} - \frac{R_{\rm f}}{\tan{\thetaf}}
    \label{eqn:thickness}
\end{equation}
where $h(r)$ is the thickness of the droplet given as a function of the distance from the center $r = 0$, $R_{\rm f}$ and $\theta_{\rm f}$ are fitting parameters. The fit of the profile measured experimentally by equation (\ref{eqn:thickness}) is shown by the dashed lines in figure \ref{fig:thickness}(b) and is found to describe reasonably well the measured profiles. However, the experimental profile also exhibits a slightly squished droplet compared to a spherical cap. This is likely due to some flattening of the drop induced by inertia at impact and to some effects of gravity in the final state. The final deposit radius $R_{\rm f}$, which is a fitting parameter, is found to only differ slightly by less than $5\%$ from the direct measurement $R$ shown in figure \ref{fig:Spreading_d}(a). The contact angle for the different volume fractions is in the range $ 5\degree \leq \theta_{\rm f} \leq 15\degree$, and increases linearly with $\phi$. This contact angle range is expected for a hydrophilic substrate, and the slight increase can be due to the presence of particles. In addition, these profiles only represent one realization of the experiment. We can thus consider that equation (\ref{eqn:thickness}) works well to describe the steady-state thickness, except at the edge of the droplet, in this limit of small Weber numbers. We also observe in \ref{fig:thickness}(b) that as $\phi$ increases, the final profile $h$ of the droplet increases in the semi-dilute and dense regimes. This is consistent with what we observe in \S \ref{subsec:radial_spread}, where the radius of the drop decreases with increasing $\phi$. 

\smallskip

In figure \ref{fig:thickness}(c), we show the rescaled mean thickness profiles $h/\hmax$ over the rescaled droplet radius $r/R_{\rm f}$ for $\phi = 0 - 10\%$ at $\We = 135$, $190$, and $240$. We plot the mean thickness profile by averaging the thickness measured at different $\phi$ for a given $\We$. We observe that the mean thickness profiles at different $\We$ are self-similar. The profiles are fitted using Equation \ref{eqn:thickness}, indicated by the dashed black lines. Hence, we use a single Equation to describe the steady-state thickness profile of a fiber suspension drop. The maximum droplet thickness $\hmax(\phi)$ is shown in the inset of figure \ref{fig:thickness}(c). The error bars for $\hmax$ shown in the inset of Figure \ref{fig:thickness}(c) are extracted from five realizations of an experiment for a particular $\phi$, $\We$, and $A$. As expected, $\hmax$ decreases slightly in the dilute regime before increasing with $\phi$. To describe $\hmax(\phi)$, we can use the conservation of volume, $\Ro^3 \sim \hmax R^2 $, equation \ref{eqn:beta_fit}, and a Taylor expansion to obtain:
\begin{equation}
    \hmax (\phi) \propto \Ro(1+0.5\phi/\phi_{\rm c})
    \label{eqn:hmax}
\end{equation}
Equation \ref{eqn:hmax} is used to fit the measured $\hmax$ for the $A = 12$ fibers at different $\We$, and is shown by the dashed lines in figure \ref{fig:thickness}(c) inset.

\subsection{Surface coverage area of fibers deposited} 
\label{sec:coating_orientation}

\begin{figure*}[!h]
  \centerline{\includegraphics[width=0.9\textwidth]{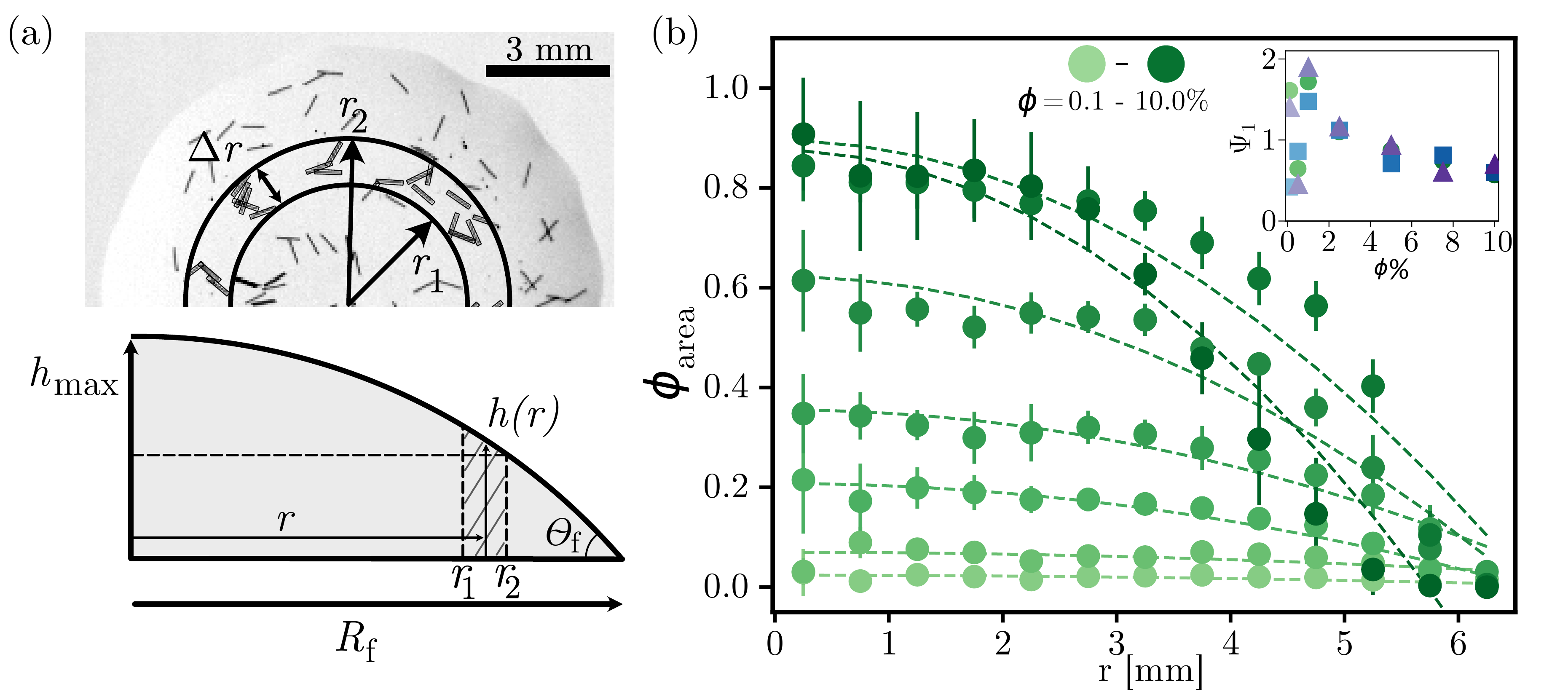}}
  \caption{(a) Top: Snapshot of a drop of fiber suspension with $A = 12$ and $\phi = 0.5\%$ impacted at $\We = 135$ at its final state. Bottom: Side-view schematic of the droplet at its final state. The shaded region represents the annulus within which the fraction of area occupied by the fibers is measured. (b) Annular area fraction $\phia$ at different radial distance $r$ from the droplet center for $A = 12$ fibers and $\phi = 0-10\%$ at $\We = 135$. The dashed lines show the fit using Equation \ref{eqn:area_fraction_final}. The fitting parameter $\Psi_{\rm 1}$ is shown in the inset for $\We = 135$ (green circles), $\We = 190$ (blue squares) and  $\We = 240$ (purple triangles).}
\label{fig:area_fraction}
\end{figure*}

 The final coating of fibers on a surface is relevant to applications from paints and coatings to surface functionalization, such as surface drag reduction \cite[]{blell2017generating, hasegawa2020microfiber}. In the top panel of figure \ref{fig:area_fraction}(a), we show a top view of a drop of fiber suspension of $A = 12$ fibers with $\phi = 0.5\%$ and $\We = 135$ at $t \simeq 100 \rm \, ms$ (\textit{i.e.}, close to its final state after impact). We use image analysis to quantify the fraction of area $\phia$ of the substrate coated by the fibers in an annular region of the droplet for various volume fractions and report the results in figure \ref{fig:area_fraction}(b) \textcolor{black}{(for further details on the procedure used to measure $\phia$, see \S S6 in the supplementary material)}. The width of the annular region considered is $\Delta r = r_{\rm 2} - r_{\rm 1} = 0.5 \rm \, mm$ and taken at various distances $r$ from the center of the droplet. The width, $\Delta r \approx L = 600 \, \rm \mu m$, and is kept constant in the present study. The volume fraction of fibers considered in figure \ref{fig:area_fraction}(b) is $\phi = 0-10\%$. In the dilute regime ($\phi \in [0, 1]\%$), $\phia$ seems almost constant over the entire range of $r$. {At first order,} the result is a homogeneous surface coating of the fibers for all the Weber numbers we investigated (see supplementary figure S7 for $\We = 190$ and $\We = 240$ for $A = 12$ fibers, figure S8 for $A = 4$ fibers). For $\phi \geq 1\%$, as $\phi$ increases, there is a gradient in $\phia$ away from the center. We expect such an observation since the droplet has the shape of a spherical cap, and the thickness of the droplet decreases away from the center (see figure \ref{fig:thickness}). This essentially results in a non-uniform coating at larger volume fractions. However, quantifying the area coated only provides a first-order description of the surface coverage by the fibers. Additional factors can be considered, such as the radial spread for different values of $\Delta r$, the overlap of the fibers, and their axial orientation within the capillary interface. Such second-order effects are unique to anisotropic particles since they can reduce the effective coated surface area. To explain our results on surface coverage, and the secondary effects unique to anisotropic particles, we first develop a model to capture $\phia$ as a function of the volume fraction $\phi$. We then use this model to interpret experimentally measured $\phia$ and to quantify the secondary effects through a fitting parameter $\Psi_{\rm 1}$.

\smallskip

The area fraction $\phia$ occupied by the fibers in the annular region is the ratio of the apparent area of the fibers detected in the annulus to the total projected area of the annulus:
\begin{equation}
    \phia = n_{\rm eff}\frac{\Psi_{\rm 1}DL}{\Delta S},
    \label{eqn:area_fraction} 
\end{equation}
where $L$ and $D$ are the length and diameter of the fibers, respectively, and $\Psi_{\rm 1}$ is a fitting factor that accounts for the apparent area detected. The projected area of the annular region, as shown in figure \ref{fig:area_fraction}(a) is $\Delta S = \pi(r_{\rm 2}^2 - r_{\rm 1}^2)$. The effective number of fibers in the annulus, $n_{\rm eff}$, can be calculated from the number of fibers per unit volume, $n$, and the volume of the annulus, $\Delta \Omega$, \textit{i.e.}, $n_{\rm eff} = n\,\Delta \Omega$. The volume in the annulus $\Delta \Omega = \Omega(r_{\rm 2}) - \Omega(r_{\rm 1})$, where $\Omega(r)$ is the volume of a section of the spherical cap, and is given by:
\begin{equation*}
    \Omega(r) = \pi r^2 h(r) + \pi \left[\hmax - h(r)\right]\,\left[3r^2 + (\hmax - h(r))^2\right]
\end{equation*}
$h(r)$ is given by Equation \ref{eqn:thickness}. We also assume that the volume of liquid lost due to splashing is negligible for the parameters considered here.

\smallskip

Now that we have an expression for $n_{\rm eff}$, we want to relate $\phia$ to the volume fraction $\phi$. By definition, the volume fraction is:
\begin{equation}
    \phi \approx N\frac{\Omega_{\rm fiber}\rho_{\rm fiber}}{\Omega_{\rm drop}\rho_{\rm drop}}
\end{equation}
where $N$ is the number of fibers in a drop, $\Omega_{\rm fiber} = \pi D^2L/4$ is the volume of a single fiber, $\rho_{\rm fiber}$ is the density of the fiber, and $\rho_{\rm drop}$ is the density of the suspension. As noted in \S \ref{sec:exp_methods}, the fibers are neutrally buoyant in the solvent, hence $\rho_{\rm fiber} \simeq \rho_{\rm drop}$. Further, the number of fibers per unit volume can also be defined with respect to the volume of an undeformed drop, $n = N/\Omega_{\rm drop}$. This simplifies the volume fraction to:
\begin{equation}
    \phi \approx n\frac{\pi D^2 L}{4}
    \label{eqn:fiber_fraction} 
\end{equation}

From equations (\ref{eqn:area_fraction}) and (\ref{eqn:fiber_fraction}), and using the fact that $n = n_{\rm eff}/\Delta \Omega$, we get the expression for $\phia$ in terms of the volume fraction $\phi$ as:
\begin{equation}
    \phia = \frac{4\,\Psii}{\pi D} \frac{\Delta \Omega} { \Delta S} \phi
    \label{eqn:area_fraction_almost_final}
\end{equation}

{As noted before, $\Psii$ is a fitting factor to account for the apparent area of the fibers captured using experiments. This factor provides second-order information on the area coated by the fibers.} In addition to $\Psii$, we consider an additional term $\Psi_{\rm 2}$ to equation (\ref{eqn:area_fraction_almost_final}), which accounts for the effect of the \textcolor{black}{squishing of the drop after impact due to both inertia and gravity. This results in a drop that is not a perfect spherical-cap, which is also noticeable as deformations at the edge in figure \ref{fig:thickness} (b). This deformation is not well represented by equation \ref{eqn:area_fraction_almost_final}, has been discussed further in \S S9 of the supplementary material.} This provides the final expression for the area fraction in terms of volume fraction:
\begin{equation}
    \phia =  \frac{4\Psii}{\pi D} \frac{\Delta \Omega} { \Delta S} \phi + \Psi_{\rm 2} 
    \label{eqn:area_fraction_final}
\end{equation}
An interesting observation from this model is that $\phia$ is independent of the fiber length, $L$, \textcolor{black}{for neutrally buoyant suspensions}. This is not directly observable from experiments. We fit the model given by equation (\ref{eqn:area_fraction_final}) to our experimental data shown in figure \ref{fig:area_fraction}(b) and observe a reasonably good fit for our results. \textcolor{black}{Further, the fitting parameter $\Psi2$ measured for the $A = 12$ fibers are provided in figure S9 (b) in the supplementary material.}

\smallskip

The values of $\Psii$ evaluated from $\phia$ for various $\phi$ at $\We = 135$, $190$, and $240$ are shown in the inset of figure \ref{fig:area_fraction}(b). We observe that $\Psii$ increases in the dilute regime but gradually asymptotes to a constant in the semi-dilute and dense regimes. The parameter $\Psii$ accounts for second-order effects introduced by the presence of fibers due to multiple factors such as overlapping, axial orientation for the fiber within the droplet, or the length of the region $\Delta r$ considered. Factors such as overlapping and orientation of the fiber with respect to the vertical axis essentially result in less of the fiber area exposed. Additional factors, such as the size of the annular region $\Delta r$, could also introduce second-order effects. For instance, although the resulting coating is uniform at low volume fractions, certain regions of $\Delta r$ may lack fibers entirely. This is reflected in the measured values of $\Psii$ for $0.1\% \leq \phi \leq 1\%$, which vary within the range $\Psii \in \left[0.5, 2 \right]$. For $\phi \geq 2.5\%$, the area covered by the fibers spans the entire droplet, eliminating non-uniformity across $\Delta r$. However, capillary effects induced by the droplet’s shape lead to a weak decrease in $\Psii$, and results in a non-uniform coating of the fibers. To investigate this parameter further, we conducted additional experiments using fibers with $A = 4$, as reported in figure S8 in the supplementary material. Preliminary observations of $\Psii$ suggest it may be a non-linear function of multiple parameters, including $\phi$, $\Delta r$, $A$, and capillary effects associated with the pinning of the contact line at the droplet boundary. Hence, further investigations are required to fully understand this parameter.

\medskip

\begin{figure*}[!ht]
    \centerline{\includegraphics[width=0.96\textwidth]{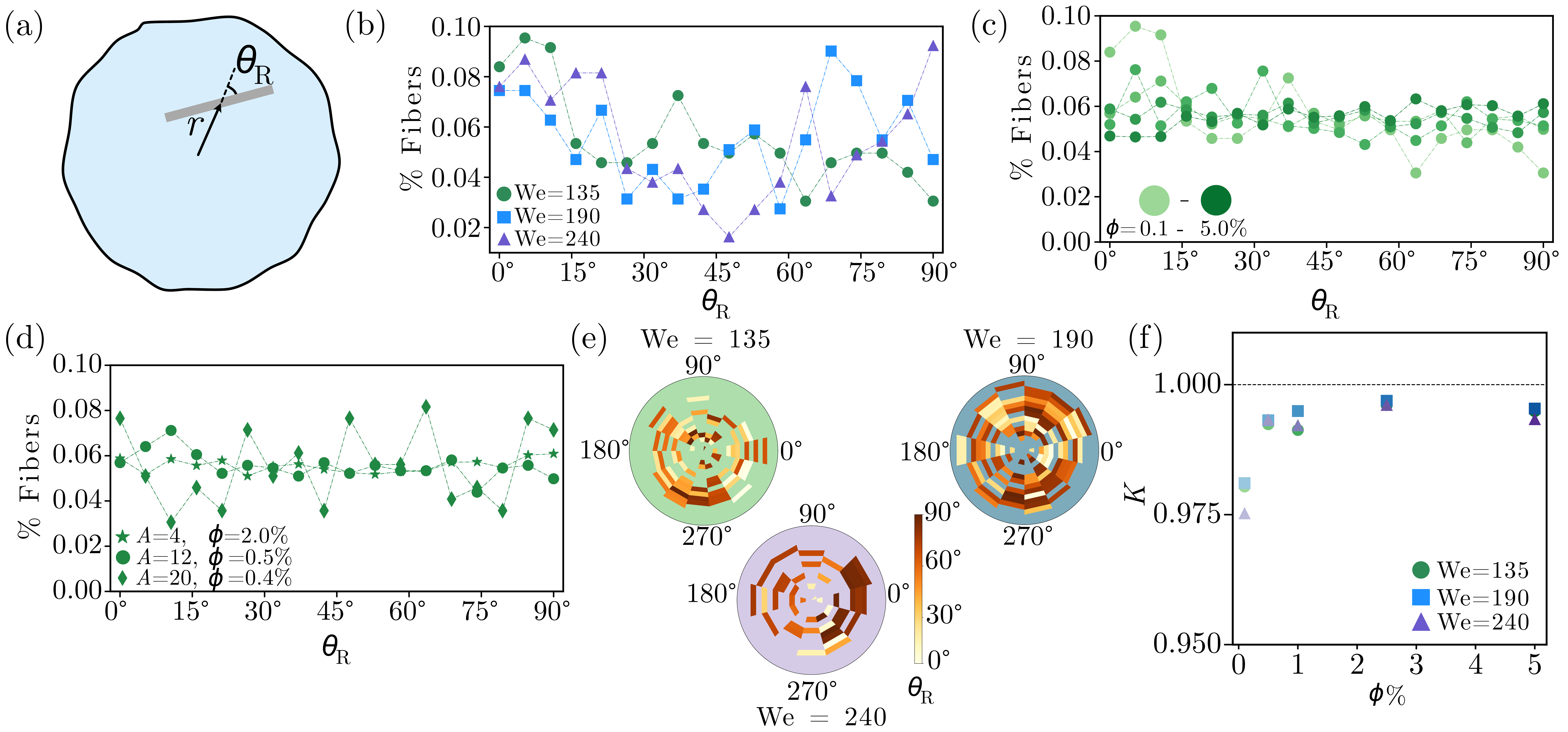}}
  \caption{(a) Schematic of a droplet with a single fiber. The radial orientation is defined as $\thetaR$, the angle the fiber makes with respect to the radial direction $r$ through the center of the fiber. (b) Distribution of $\thetaR$ for $A = 12$, $\phi = 0.1\%$ fibers at $\We = $ 135, 190, and 240. (c) Distribution of $\thetaR$ for $\phi \in \left[ 0.1, 5 \right] \%$ of $A = 12$ fibers at $\We = 135$ (d) Distribution of $\thetaR$ for $A = 4, \phi = 2\%$ (star), $A = 12, \phi = 0.5\%$ (circle), $A = 20, \phi = 0.4\%$ (diamond) at $\We = 135$. The dotted lines in figures (b)-(d) are only guides for the eye. (e) Heat map describing the radial spread of $\thetaR$ for $A = 12$, $\phi = 0.1\%$ at $\We = 135, 190,$ and $240$, corresponding to green, blue, and purple background, respectively. (f) Isotropy factor $K$ for different values of $\phi$ for $A = 12$ fibers.}
\label{fig:orientation}
\end{figure*}


\subsection{{Final orientation of the fiber}}
\label{sec:fiber_orientation}

\smallskip

{Capillary dynamics can play a crucial role in influencing fiber orientation during droplet deposition, as the balance of surface tension, inertia, and viscous dissipation aligns fibers in the suspension \cite{Deegan1997, yarin2006}. This alignment critically affects the mechanical, thermal, and functional properties of the resulting material, making it relevant to applications in industries such as additive manufacturing and biomedical engineering \cite{Jawaid2018, pilato2013advanced}.} Further, the radial orientation is also required to calculate the deviatoric stresses in suspensions of anisotropic particles (see, e.g., the work by Shaqfeh and Fredrickson \cite{shaqfeh1990hydrodynamic} for more details).

\smallskip


In the present study, we can directly measure the radial orientation $\thetaR$ in the dilute and semi-dilute regimes, where overlapping fibers remain distinguishable. We define $\thetaR$ as the angle formed between the fiber and the radial vector $r$ passing through its center, as illustrated in figure \ref{fig:orientation}(a). In figure \ref{fig:orientation}(b)-(d), we summarize the measured values of $\thetaR$ after the droplet has reached its final state, \textit{i.e.}, $t \simeq 100 \, {\rm ms}$. We measure this orientation for $A = 12$ fibers with $\phi \in [0.1, 5]\%$, $A = 4$ fibers with $\phi = 2\%$, and $A = 20$ fibers with $\phi = 0.4\%$, at $\We = 135-240$. The procedure involves using a custom routine developed in Python that can distinguish between two overlapping fibers (see \S S10 in the supplementary material).

\smallskip

In figure \ref{fig:orientation}(b)-(d), we present the fraction of fibers detected as a function of $\thetaR$. The distribution is based on seven experimental realizations for each parameter set, resulting in a dataset that ranges from over 100 to 5,000 points, depending on the value of $\phi$. Figure \ref{fig:orientation}(b) shows the orientation measured for $\phi = 0.1\%$ of the $A = 12$ fibers at $\We = $ 135, 190, and 240. In this dilute regime, we observe a slight anisotropy in the distribution of $\thetaR$. Figure \ref{fig:orientation}(c) captures the influence of volume fraction in the range $\phi \in [0.1, 5]\%$ for the $A = 12$ fibers at $\We = 135$. Indeed, as $\phi$ increases, we observe that the distribution of $\thetaR$ becomes homogeneous, \textit{i.e.}, the fibers are increasingly randomly distributed. This could be due to an increase in fiber-fiber interactions or an increase in viscous dissipation, which minimizes the rearrangement of fibers in any preferred direction. In figure \ref{fig:orientation}(d), We probe the role of fiber geometry $A$, for aspect ratios $A = 4$, $12$, and $20$. We chose the volume fraction $\phi$ for different $A$ such that we have sufficient fibers to quantify and that the suspensions have similar enough viscosity. Once again, we see an isotropic distribution for $\thetaR (A)$. The isotropy appears to decrease weakly with $A$. This observation suggests that of the two factors driving isotropy, \textit{i.e.}, viscous dissipation and fiber-fiber interaction, the latter is likely more important. In summary, from figure \ref{fig:orientation}(b)-(d), we note that the distribution of $\thetaR$ is within a narrow range of 0 to 0.1 across different $\We$, $\phi$, and $A$. Despite the slight differences observed, the overall steady-state radial orientation is isotropic across the range of parameters we measure.

\smallskip

Figure \ref{fig:orientation}(b)-(d) provides insight into the net distribution of $\thetaR$. However, it does not reveal any information on the distribution of $\thetaR$ in the radial or angular direction. To probe this, we quantify the distribution of $\thetaR$ as a function of $r$ and $\theta$. This is shown in figure \ref{fig:orientation}(e) for a suspension with $\phi = 0.1\%$ and $\We = $ 135, 190, and 240. The heat map provides information on radial or angular preference for the distribution of $\thetaR$, if any. However, figure \ref{fig:orientation}(e) does not reveal any such preferential orientation. We observe this to be true for all the $\We$ across $\phi \in \left[0.1, 5\right]\%$ for the $A = 12$ fibers.

To summarize the radial orientations measured in figure \ref{fig:orientation}(b)-(d), we introduce an isotropy factor $K$. At a particular $\phi$, $\We$, and $A$, we define it as $K = 1-\sigma$, where $\sigma$ is the standard deviation of the distribution of $\thetaR$. Hence, $K = 1$ here represents an isotropic distribution. We quantify the isotropy $K$ as a function of $\phi$ in figure \ref{fig:orientation}(f). The fibers considered are $A = 12$  with $\phi \in \left[0.1, 5\right]\%$ at $\We =$ 135, 190, and 240. As $\phi$ increases, $K$ converges to 1. Hence, despite the narrow range $K$, we observe that as the volume fraction increases, the randomness in the distribution converges to $1$. Similarly, we report the influence of fiber geometry $A$ on the measured isotropy in figure S11 in the supplementary material. In summary, the quantification of $K$ confirms that for the range of $\phi$, $\We$, and $A$ considered in this study, the deposited fibers in the suspension exhibit weak to no preferential orientation. We mention here that during the dynamic spreading of the droplet, there is some preferential alignment for the droplet, \textcolor{black}{initially in the radial direction, and perpendicular to the radius when the liquid forms a crest at the edge of the droplet} (see Movie 1 and Movie 2). However, \textcolor{black}{this orientation is simply the alignment of the fibers during the spreading}, and as the spreading approaches steady-state, the $\thetaR$ distribution becomes isotropic.

\medskip

\section{{Conclusion}}
\label{sec:conclusions}

The present study investigated the impact of a drop of fiber suspension on a hydrophilic substrate. The parameters explored include fibers of different volume fractions $\phi$ spanning the dilute, semi-dilute, and dense regimes. The droplets were released from various heights, resulting in different $\We$ and $\Reynolds$. While most of the experiments used fibers with an aspect ratio $A = 12$, a few additional experiments were conducted with $A = 4$ and $A = 20$ fibers to extend the broad parameter space.

\smallskip

We quantified several key features typically considered in studies on droplet impact: the spreading dynamics and maximum radius of the droplet \cite{nicolas2005spreading, grishaev2015complex}, the steady-state thickness of the drop \cite{raux2020spreading}, the resulting coating of the fibers \cite[][]{xie2025effect, maddox2024capillary}, and the steady-state radial orientation \cite{ jeong2023deposition, pashazadeh2024predicting, blell2017generating}. For our investigations on spreading dynamics and final droplet size, we hypothesized that as the volume fraction increases, the resulting size of the drop decreases due to the increase in apparent viscosity. This is indeed seen in our results for the fibers in semi-dilute and dense regimes. Further, the decrease in droplet radius implies that the droplet thickness $\hmax$ should increase with $\phi$. We verified this by implementing an image spectroscopy technique for measuring the thickness of fiber suspensions.

\smallskip

Interfacial effects play a crucial role in determining the surface area coated by particles during droplet deposition, for instance, \textcolor{black}{by controlling the release of particles during extrusion or spreading of gels used in self-healing materials \cite{hia2016electrosprayed, truby2016printing},} or by driving the flow of particles toward the contact line through capillary flow during the evaporation of colloidal suspensions \cite{hu2002evaporation}. This phenomenon, influenced by particle size, surface tension, and drying dynamics, is critical in applications like printing, coatings, and biomedical devices \cite{Deegan1997, Dufresne2003}. To analyze the fiber coating, we define $\phia$ as the surface area coated by the fibers. The analysis revealed that at low $\phi$, the resulting coating seems uniform at first order, while at higher $\phi$, it becomes inhomogeneous. Predicting the surface coating without experiments is challenging due to the anisotropic nature of the particles. A key difference between spherical particles and fibers is that fibers can overlap, especially at larger $A$. This introduces non-linear secondary effects that depend on multiple parameters such as the aspect ratio $A$, the volume fraction $\phi$, and the annular region $\Delta r$ considered. A description the coated area is essential to improve predictive capabilities for additive manufacturing or spray coating applications involving capillary deposition \cite{truby2016printing, turton2005scale}, since the surface area coated is critical to the strength of the material. {We propose a model to capture} the evolution of the coated area, $\phia(r)$, which includes a fitting parameter, $\Psi_{\rm 1}$, to describe the second-order effects introduced by the fibers.

Capillary interfaces influence fiber orientation due to flow dynamics generated by the balance of surface tension, inertia, and viscous dissipation, resulting in fiber alignment during wetting, drying, and deposition processes \cite{Deegan1997, yarin2006}. The resulting fiber orientation critically influences the mechanical, thermal, and functional properties of materials, enhancing strength, conductivity, and porous media flows in industries like aerospace, automotive, and biomedical engineering \cite{Jawaid2018, pilato2013advanced, yusoff2023optimization, wang2021wireless}. Applications range from improving structural properties in 3D-printed composites, filtration systems, energy efficiency, and tissue scaffolds through precise alignment strategies \cite[See Refs.][]{regalla2020strength, yusoff2023optimization, Koffler2019, haase2017multifunctional}. Our measurements of the radial orientation, $\thetaR$, reveal that despite the different $\phi$, $\We$, and $A$ explored, the resulting $\thetaR$ is isotropic. Hence, for the range of parameters considered here, the droplet impact process does not introduce any preferential orientation. Such an observation is useful in designing processes where uniform spreading of the fibers is necessary. The image analysis routine we developed to measure the orientations improves upon previous approaches by being able to distinguish between overlapping fibers with reasonable accuracy \cite{jeong2023deposition, pashazadeh2024predicting}. While the presented routine has its limitations arising from the accuracy of the segmentation routine used, introducing such an approach is crucial to measure $\thetaR$ at higher volume fractions.

\smallskip

Characterizing the capillary dynamics of suspension flow is complex but important for practical coating and industrial applications. The fiber suspensions considered here introduce additional complexity compared to spherical particles. Further studies are required to control the orientation of the fibers on the substrate, for instance, by laterally impacting suspension droplets. Nevertheless, we revealed key insights into the peculiar capillary dynamics of fiber suspensions that could aid in predictions for industrial processes. While the present study aims to discuss some first results on the drop impact of fiber suspensions, many future directions remain to be explored as capillary flows of fiber suspensions have not yet been as studied as well as capillary flows of spherical particle suspensions. In this work, investigations on the role of the aspect ratio $A$ were limited. For studying the final size of the droplet, further experiments with different aspect ratios would refine the model for the fitting parameter $\alpha$ as a function of $A$. Similar studies could explore the role of $A$ on droplet thickness $h$ using the image spectroscopy technique we presented. Additionally, the technique can resolve dynamic droplet thickness and quantify the role of fibers in modifying capillary waves, such as during droplet spreading. Further investigations are also needed to probe the role of substrate wettability and orientation, which may be critical for designing 3D scaffolds or targeted drug delivery using droplet extrusion. Another important direction is to investigate if one could introduce a preferential orientation for such fibers deposited by drop impact, for instance, by controlling the orientation of the substrate or the nozzle or by freezing the droplet during the spreading stage to arrest the dynamic orientation of the fibers in motion.

\bigskip\bigskip

\section*{Declaration of Competing Interest}

The authors declare that they have no known competing financial interests or personal relationships that could have appeared to influence the work reported in this paper.

\section*{Acknowledgments}

This material is based upon work supported by the National Science Foundation under NSF CAREER Program Award CBET Grant No. 1944844.

\bibliographystyle{main}
\bibliography{biblio_drop}

\end{document}


\maketitle

\vspace{5mm}

\Large {{\textbf{Movies of the experiments}}}

\normalsize 

\vspace{5mm}

\noindent - Movie 1 corresponds to the impact of $\phi = 1\%$ suspension droplet with $A = 12$ fibers at $\rm We = 190$.\\
\smallskip
\noindent - Movie 2 corresponds to the impact of $\phi = 10\%$ suspension droplet with $A = 12$ fibers at $\rm We = 190$.

\clearpage


\section{Droplet free fall and impact}

\begin{figure}[h!]
 \begin{center}
 \includegraphics[width = 0.75 \textwidth]{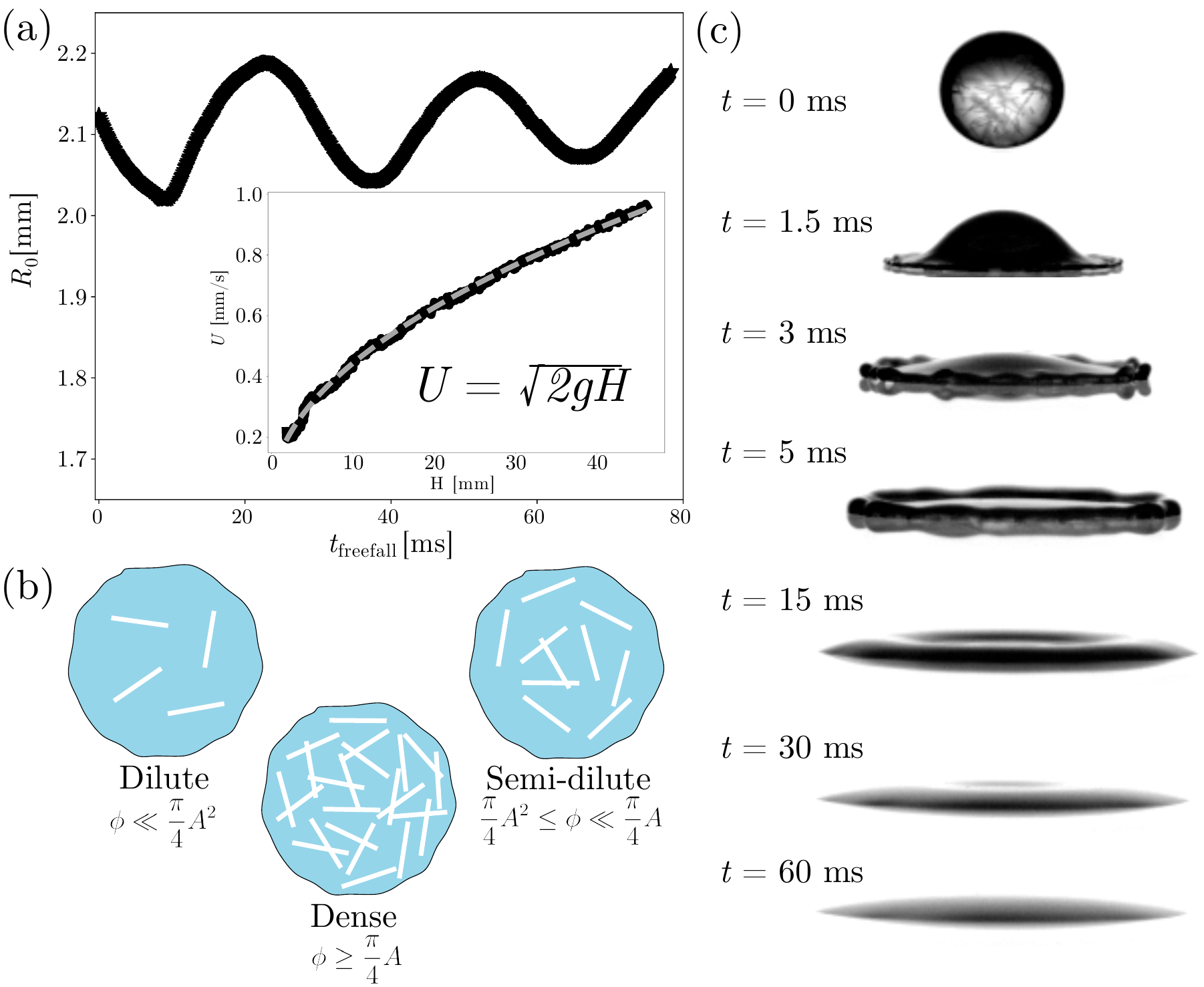}
 \caption{(a) Time-evolution of the radius $R_{\rm 0}$ of the droplet during free fall. Inset: free fall droplet velocity $U$ as a function of the height $H$. (b) Schematic showing the separation between the dilute, semi-dilute, and dense regimes. (c) Side-view of a time sequence of the impact and spreading of a fiber suspension droplet for $A = 12$ and $\phi = 0.5\%$.}
 \label{fgr:SM_Fig1}
  \end{center}
\end{figure}

Figure \ref{fgr:SM_Fig1}(a) reports the time evolution of the radius of the droplet $R_{\rm 0}(t)$ during its free fall before it impacts the substrate. After pinch-off, the radius of the droplet oscillates about a mean value due to the balance between capillarity and inertia \cite{wang2018unsteady}. The inset in figure S1(a) shows the droplet velocity in free fall as a function of the height $H$. 

In figure \ref{fgr:SM_Fig1}(b), we show schematically the demarcation between the dilute, semi-dilute, and dense regimes. The demarcation between the different regimes as a function of volume fraction $\phi$ and the aspect ratio $A$ for a neutrally buoyant fiber suspension is obtained following the definition by Butler \textit{et al.} \cite{butler2018microstructural}. Using the number density, $n = N/\Omega_{\rm drop}$, where $N$ is the number of fibers in a volume $\Omega_{\rm drop}$, then:

\[\setlength\arraycolsep{1pt}
    \begin{array}{rcccl}
                & ~ & n & \ll & 1/L^3 \qquad \,\,\,\,\,\text{for dilute suspensions}\\
        1/L^{3} & \leq & n & < &  1/DL^{2} \qquad \text{for semi-dilute suspensions}\\
                & ~ & n & \geq &  1/DL^{2} \qquad \text{for dense suspensions}
    \end{array}
\]

In the present work, we use fibers of diameters $D$, length $L$, and thus of volume $\Omega_{\rm fiber} = \pi D^2L/4$. The volume fraction is then $\phi = (N\,\Omega_{\rm fiber}\,\rho_{\rm fiber})/(\Omega_{\rm drop}\,\rho_{\rm drop}) = n\,\Omega_{\rm fiber}$, since the particles are neutrally buoyant. As a result, the number of fibers per unit volume in the drop is  

\begin{equation}
  n = \frac{4\phi}{\pi D^{\rm 2}L}
\end{equation}

Substituting this expression into the limits defined by Butler \textit{et al.} \cite{butler2018microstructural} for fibers with aspect ratio $A = L/D$, we have: 

\[\setlength\arraycolsep{1pt}
    \begin{array}{rcccl}
                 & ~ & \phi & \ll & \frac{\displaystyle \pi}{\displaystyle 4A^2} \quad \,\,\text{for dilute suspensions}\\[3mm]
\frac{\displaystyle \pi}{\displaystyle 4}A^2 & \leq & \phi & < &  \frac{\displaystyle \pi}{\displaystyle 4\,A} \qquad \text{for semi-dilute suspensions}\\[3mm]
                & ~ & \phi & \geq &  \frac{\displaystyle \pi}{\displaystyle 4\,A} \qquad \text{for dense suspensions}
    \end{array}
\]

For $A = 4$, $A = 12$, and $A = 20$, the range of dilute, semi-dilute, and dense regimes is reported in table \ref{tab:fiber_prop}:

\smallskip

\begin{table}[!h]
  \begin{center}
\def~{\hphantom{0}}
  \begin{tabular}{lccccc}
        \smallskip
        $A$  & $L \, \rm [\mu m]$ & $D \, \rm [\mu m]$ & $\phi_{\rm dilute}$ & $\phi_{\rm semi-dilute}$ & $\phi_{\rm dense}$ \\[3pt]
        \hline
        \smallskip
        4   &     200            &        50 &    $\leq$ 5\%         &  5-20\%           &  $\geq$ 20\% \\
        \smallskip
        12  &     600            &        50 &   $\leq$ 0.54\%       &  0.54-6.5\%       &  $\geq$ 6.5\%\\
        \smallskip
        20  &     1000           &        50 &   $\leq$ 0.20\%       &  0.20-4\%         &  $\geq$ 4\%\\
  \end{tabular}
  \caption{Limits of volume fraction $\phi$ between the dilute, semi-dilute, and dense regimes for fiber suspensions of aspect ratios $A = $ 4, 12, and 20.}
  \label{tab:fiber_prop}
  \end{center}
\end{table}

In figure \ref{fgr:SM_Fig1}(c) an example of a side view of the impact and spreading of a droplet of fiber suspension with $A = 12$ and $\phi = 0.5\%$ on a hydrophilic substrate.


\newpage
\section{Size distribution}

\begin{figure}[h!]
 \begin{center}
 \includegraphics[width = 0.75\textwidth]{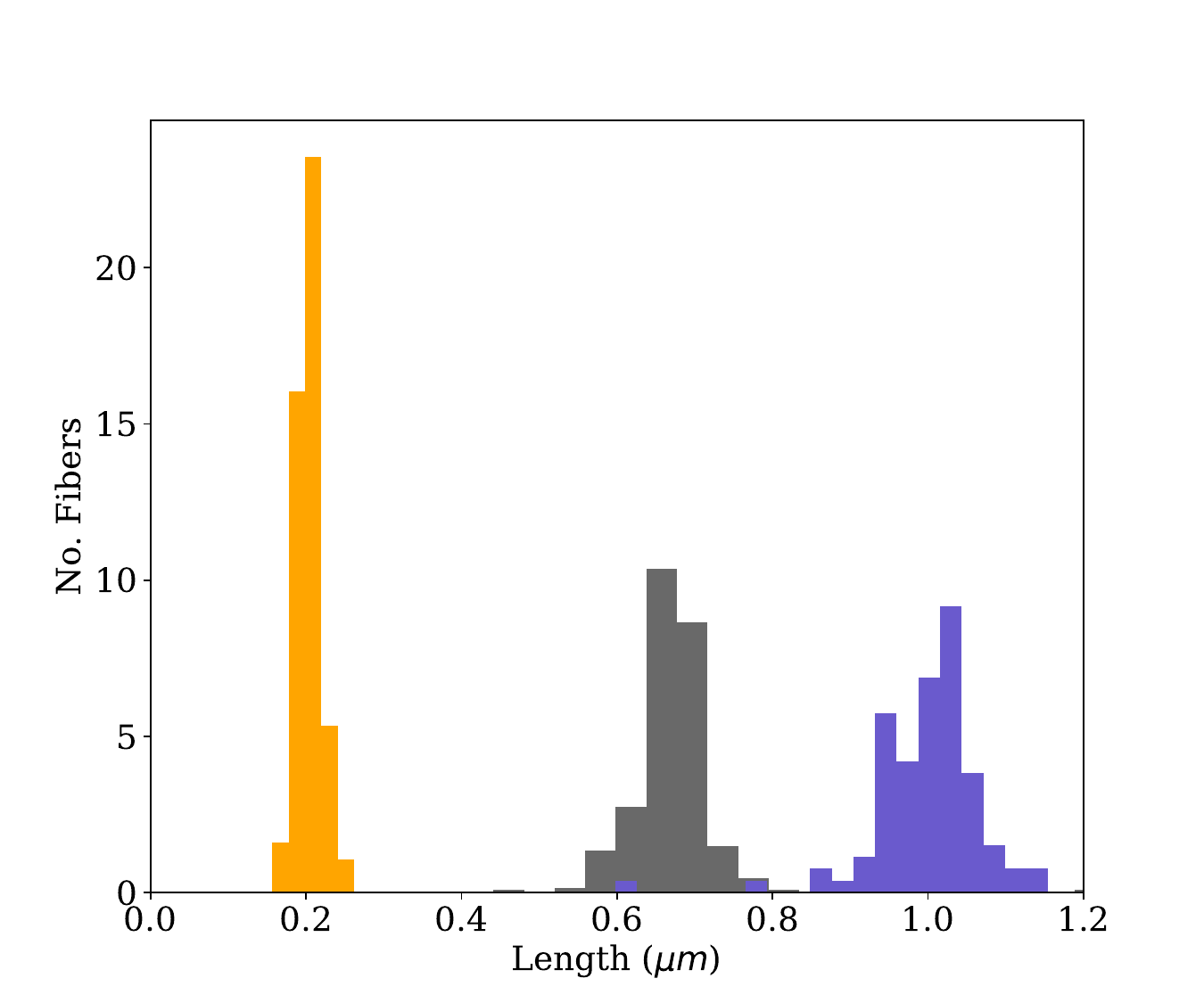}
 \caption{Size distribution measured for the length $L$ of the $A = 4$ (orange bars), $A=12$ (grey bars), and $A=20$ (purple bars) fibers.}
 \label{fgr:SM_Fig2}
  \end{center}
\end{figure}

The size distribution, measured with image analysis, for the fibers of different aspect ratios used in this study is reported in figure \ref{fgr:SM_Fig2}. The fiber length for the $A = 4$, $12$, and $20$ is measured in the range $L = 200 \pm 40 \, \rm \mu m$, $L = 660 \pm 80 \, \rm \mu m$, and $L = 1000 \pm 100 \, \rm \mu m$, respectively.

\clearpage

\section{Volume fraction post-extrusion }

\begin{figure}[h!]
 \begin{center}
 \includegraphics[width = 0.6\textwidth]{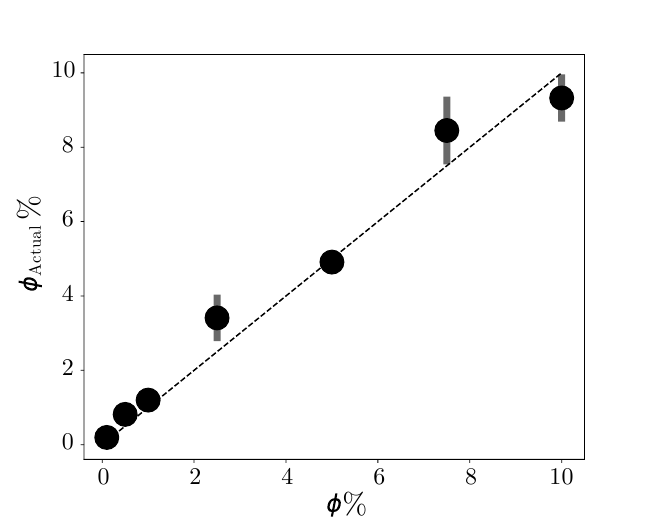}
 \caption{Comparison of the volume fraction of the prepared suspension $\phi$ and the actual volume fraction of the extruded drop $\phi_{\rm actual}$. The dotted line shows $\phi=\phi_{\rm actual}$, \textit{i.e.}, no self-filtration.}
 \label{fgr:SM_Fig3}
  \end{center}
\end{figure}

For the range of volume fractions and aspect ratios investigated in the study, we do not observe any self-filtration, \textit{i.e.}, an extruded volume fraction smaller than the one in the syringe \cite{haw2004jamming}. We report in figure \ref{fgr:SM_Fig3} experimental measurements for $A = 12$ fibers and $0.1\% \leq \phi \leq 10\%$. The experimental procedure involved extruding droplets from a syringe in which the volume fraction is known onto a container and measuring the mass of the extruded suspension droplet before and after removing the solvent. The solvent, which is a 40\%/60\% by weight mix of water/glycerol, was removed by washing away the glycerol with water. The resulting water and fiber mixture was then dried, and the mass of the fibers was measured. 

\newpage


\section{Spreading factor for $A = 4$ and $A = 20$ fibers}

\begin{figure}[h!]
 \begin{center}
 \includegraphics[width = 1\textwidth]{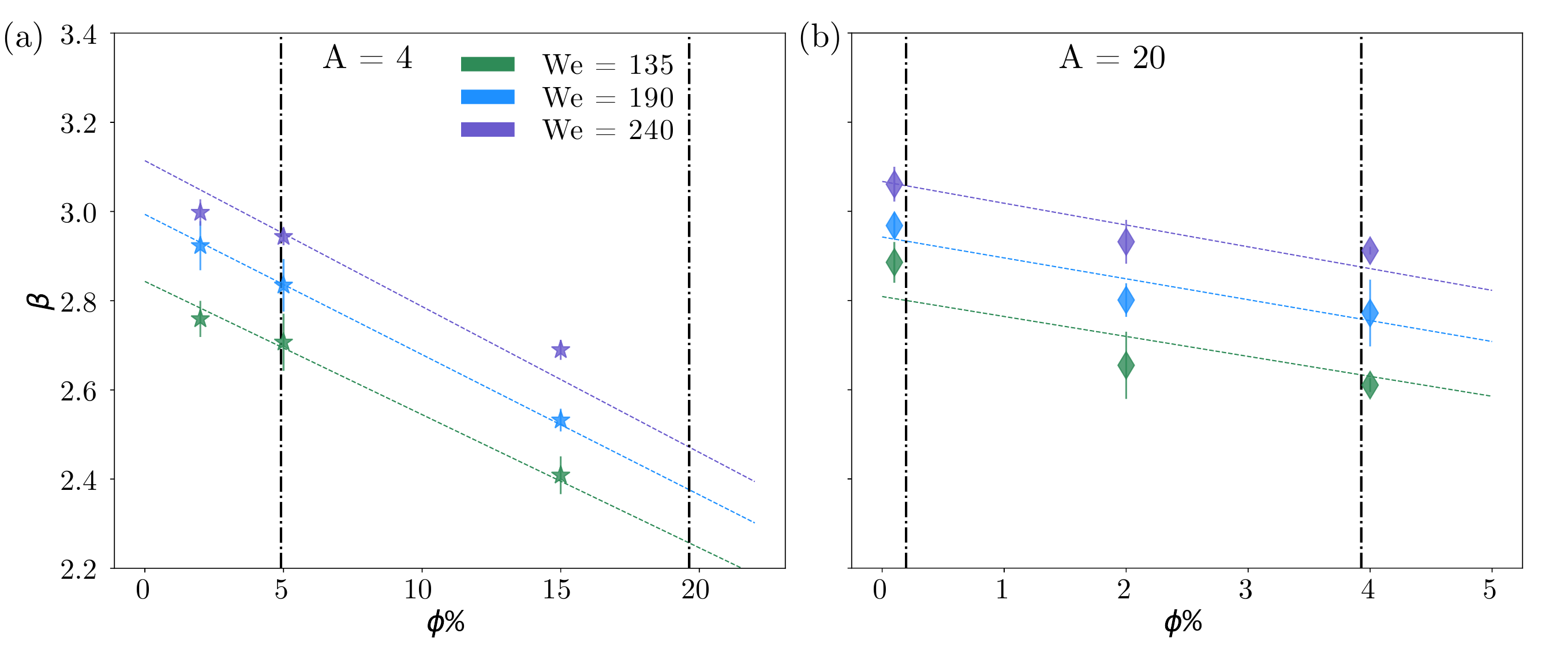}
 \caption{Spreading factor $\beta(\phi)$ for suspension droplets with fibers of aspect ratio (a) $A = 4$, and (b) $A = 20$. The experimental data is fitted with equation 4 (shown by the dashed lines) from the main article. The vertical dashed-dotted line separates the dilute, semi-dilute, and dense regimes.}
 \label{fgr:SM_Fig4}
  \end{center}
\end{figure}

\vspace{5mm}
The spreading factor $\beta(\phi)$ measured for the $A = 4$ and $A = 20$ fibers is shown in figure \ref{fgr:SM_Fig4}(a) and \ref{fgr:SM_Fig4}(b), respectively. The dashed lines are fits based on equation 4 from the main article. The volume fraction considered for the $A = 4$ fibers, shown in figure \ref{fgr:SM_Fig4}(a), are $\phi = $ 2, 5 and 15\%, which gives a fitting factor $\alpha = 0.93 \pm 0.02$. For the $A = 20$ fibers, shown in figure \ref{fgr:SM_Fig4}(b), $\phi = $ 0.1, 2, and 4\%, and $\alpha = 0.99 \pm 0.02$. For $A = 4$ and $A = 20$ fibers, we consider $\phi_{\rm c} = 47.6\%$, and $\phi_{\rm c} = 31.4\%$ respectively, based on the investigations reported by Bounoua \textit{et al.} \cite{bounoua2019shear}.

\newpage


\section{Measurement of the droplet thickness}

\vspace{5mm}

\begin{figure}[h!]
 \begin{center}
 \includegraphics[width = 1\textwidth]{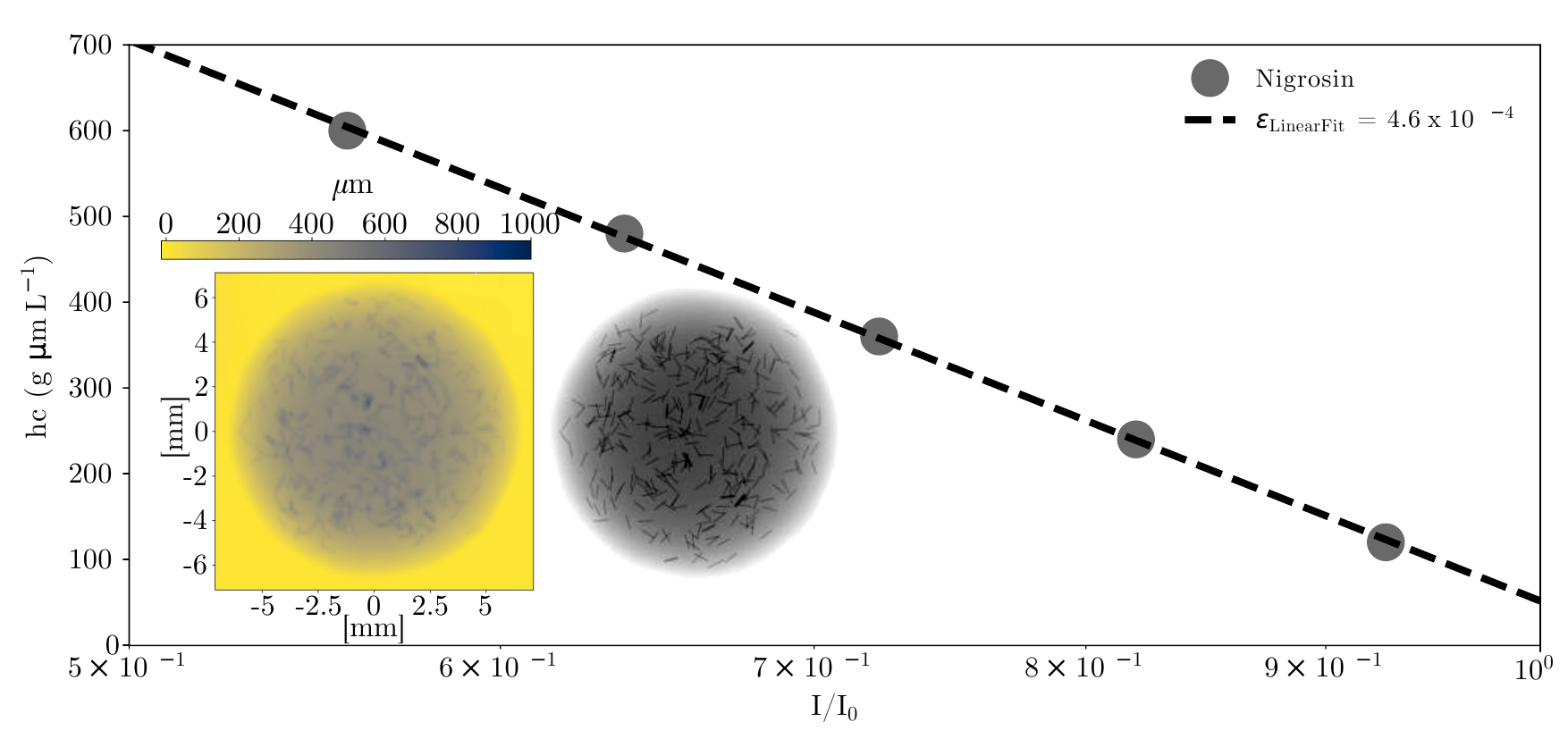}
 \caption{Calibration plot for the thickness measurements. The solid points are experimentally measured intensity $I/I_{\rm 0}$. The dashed lines are the fit using equation \ref{eqn:beer_lambert}. Inset: raw (right) and computed (left) image of the droplet after its impact on the substrate.}
 \label{fgr:SM_Fig5}
  \end{center}
\end{figure}

We measure the thickness of the droplet by using a light absorption technique reported by Vernay \textit{et al.} in a different configuration \cite{vernay2015free}. We prepare two sets of suspensions for the experiments. For the set of experiments with dye, we prepare the suspension by adding the fibers to a solvent made of 40\%/60\% by weight glycerol with $1.2 \rm \, g\,L^{-1}$ of Nigrosin (Sigma-Aldrich). Note that the Nigrosin does not modify the surface tension of the liquid. To calibrate the evolution of the absorption versus thickness, we use capillary tubes (VitroCom) of known thickness in the range $h = $ 100 - 500 $\rm \mu m$ and fill them with the solvent. We measure the intensity $I$ of the solvent inside the capillary tube and the reference intensity $I_{0}$ for an empty tube. The normalized intensity $I/I_{0}$ measured is plotted for various capillary tube thicknesses as $h\,c$ in figure \ref{fgr:SM_Fig5}(a). The calibration data is fitted with a modified form of the Beer-Lambert law:
\begin{equation}
\label{eqn:beer_lambert}
    \log(I_{\rm 0}/I) = \varepsilon_{\rm m}hc + c_{\rm i}
\end{equation}
where $\varepsilon_{\rm m}$ is the absorptivity, and $c_{\rm i}$ is constant, both being fitting parameters. We use this calibration in the high-speed recording to extract the thickness of the drop for $t > t_{\rm coll}$ as it spreads on the substrate. The right inset in figure \ref{fgr:SM_Fig5} shows an actual image of a drop of suspension with $A = 12$ and $\phi = 1\%$ at $t \simeq 100 \rm \, ms$. In the left inset in figure \ref{fgr:SM_Fig5}, we show the corresponding thickness profile extracted from the absorption technique. Using the thickness analysis routine, we can extract the thickness profiles for all the fiber fractions we investigate. The thickness profile is extracted by averaging the mean intensity over an area of $3\times3 \rm \, pixel^2$ for computational efficiency. The fibers in the drop are highlighted by darker regions compared to the surrounding liquid. 

\newpage


\section{Procedure for Quantifying Area Fraction}

\begin{figure}[h!]
 \begin{center}
 \includegraphics[width = 1\textwidth]{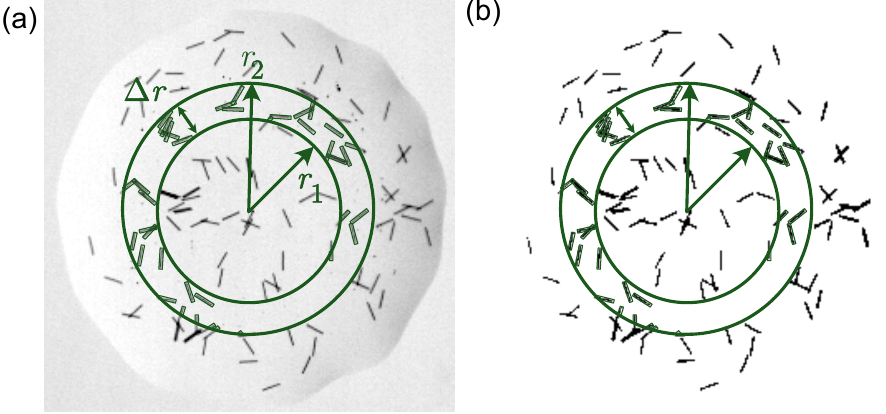}
 \caption{Quantifying the area fraction for the $A = 12$ suspensions with $\phi = 0.1\%$ at $\rm We = 135$. (a) Image of the droplet at $t = 60 \, \rm ms$, and (b) the corresponding binarized image with the liquid removed.}
 \label{fgr:SM_Fig6}
  \end{center}
\end{figure}

The method adopted to quantify the area fraction here consists of binarizing the image of the droplet at steady-state such that we threshold the liquid out of the image. We do this by using the set of experiments for which the nigrosin dye was not added to the solvent. This results in a post-binarization image where only the fibers are visible. The area fraction $\phi_{\rm area}$ is then quantified by measuring the area occupied by the black pixels, which essentially represents the 2D projected area of the fibers, within an annular region $\Delta r$ as shown in figure \ref{fgr:SM_Fig6}. This presents a simple routine that can be implemented in any live measurement application involving coating of fibers.

\newpage


\section{Area Fraction for $A = 12$ and $0.1\% \leq \phi \leq  10\%$ at $\rm We = $ 190 and 240}

\begin{figure}[h!]
 \begin{center}
 \includegraphics[width = 1\textwidth]{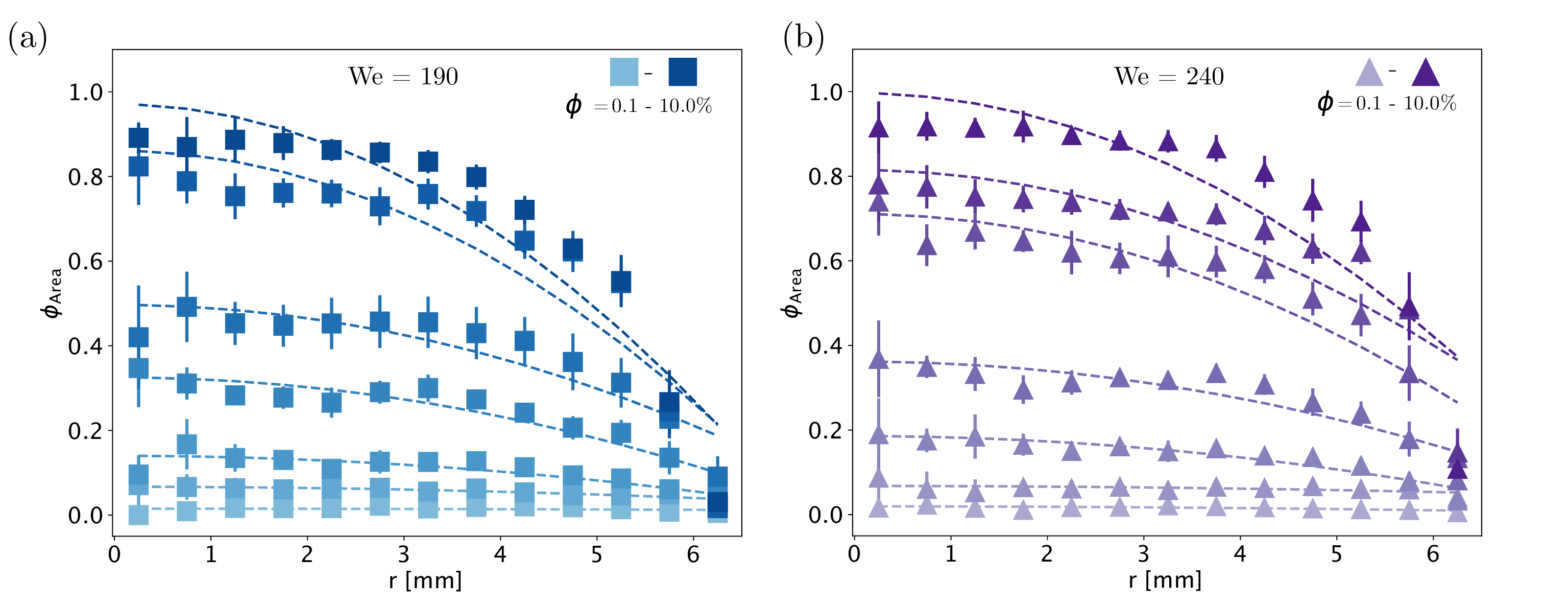}
 \caption{Evolution of the area fraction with the radius for the $A = 12$ suspensions with $0.1\% \leq \phi \leq  10\%$ at (a) $\rm We = 190$ and (b) $\rm We = 240$.}
 \label{fgr:SM_Fig7}
  \end{center}
\end{figure}

\vspace{5mm}
As presented in subsection 3.4 and figure 6 in the main article, figure \ref{fgr:SM_Fig7}(a)-\ref{fgr:SM_Fig7}(b) show the area fraction $\phi_{\rm area}(r)$ for $\rm We = 190$ and $240$, respectively. Each data point corresponds to the calculation in an annular region of width $\Delta r=0.5\,{\rm mm}$. The dashed lines represent the fit using equation 11 from the main article, with the fitting parameter $\Psi_{\rm 1}$ shown in the inset of figure 6(a) in the main article.

\newpage


\section{Area Fraction for $A = 4$ and $2\% \leq \phi\leq  15\%$}

\begin{figure}[h!]
 \begin{center}
 \includegraphics[width = 1\textwidth]{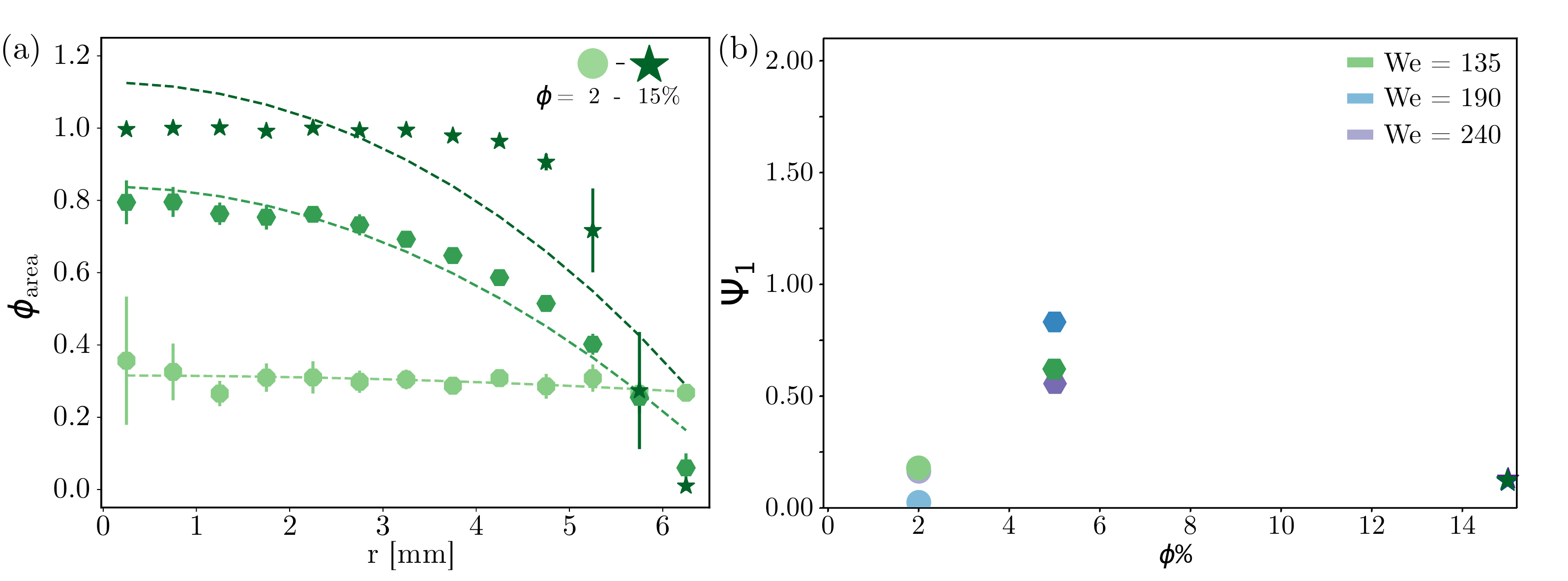}
 \caption{(a) Area fraction $\phi_{\rm area}$ for the $A = 4$ suspensions with $\phi =$ 2, 5, and 15\% at and $\rm We = 135$. (b) The corresponding fitting parameter $\Psi_{\rm 1}$ at $\rm We =$ 135, 190, and 240.}
 \label{fgr:SM_Fig8}
  \end{center}
\end{figure}

\vspace{5mm}
figure \ref{fgr:SM_Fig8} (a) reports the radial evolution of $\phi_{\rm area}(r)$ for fibers of aspect ratio $A = 4$ and volume fractions in the range $2\% \leq \phi\leq  15\%$ at a Weber number $\rm We = 135$. The dashed lines represent the fit using equation 11, with the fitting parameter $\Psi_{\rm 1}$ shown in figure \ref{fgr:SM_Fig8} (b). For $A = 4$, the values measured for $\Psi_{\rm 1}$ are much smaller compared to observations from the $A = 12$ fibers. This is an interesting observation, since even at dilute fiber concentrations, the presence of fibers likely introduces second order effects. However, further experiments with fibers of different aspect ratios are required to delineate the role of $A$ of this parameter.

\newpage


\section{Fitting parameter $\Psi_{\rm 2}$}

\begin{figure}[h!]
 \begin{center}
 \includegraphics[width = 0.99\textwidth]{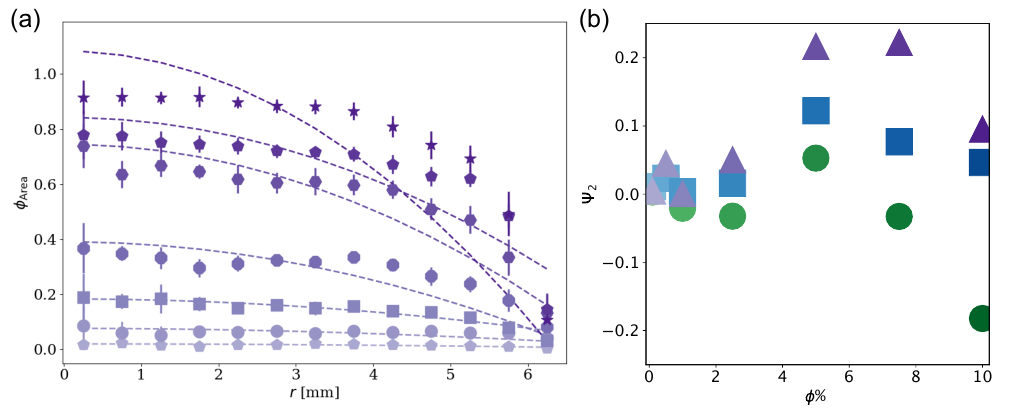}
 \caption{(a) Evolution of the area fraction for the $A = 12$ suspensions for $\phi \in [0.1, 10]\%$ at $\rm We = 240$ when $\Psi_{\rm 2}$ is not used. (b) Experimentally measured values of $\Psi_{\rm 2}$ for $\phi \in [0.1, 10]\%$ at $We = 135$, $190$, and $240$.}
 \label{fgr:SM_Fig9}
  \end{center}
\end{figure}

In figure \ref{fgr:SM_Fig9} (b) below, we show the fitting parameter $\Psi_{\rm 2}$ measured for the area fraction occupied by the $A = 12$ fibers for $\phi \in [0.1, 10]\%$. We note that $\Psi_{\rm 2} \in [-0.2, 0.2]$, and weakly increases in magnitude as $\phi$ increases. A similar trend is also observed for the $A = 4$ fibers, where $\Psi_{\rm 2} \in [0, 0.4]$. equation 11 used in the main article to model $\phi_{\rm area}$ has $\Delta \Omega$, the volume of a section of a spherical cap, as the dominant term. Hence, if we fit the area fraction without accounting for $\Psi_{\rm 2}$, it results in a more spherical cap like shape for the fit, as shown in figure \ref{fgr:SM_Fig9}(a). The parameter $\Psi_{\rm 2}$ here is a correction term to equation 11, which accounts for the deformations or squishing introduced to the droplet due to inertia and gravity \cite{gennes2004capillarity}.

\newpage


\section{Orientation of the fibers on the substrate}

\smallskip

\begin{figure}[h!]
 \begin{center}
 \includegraphics[width = 0.99\textwidth]{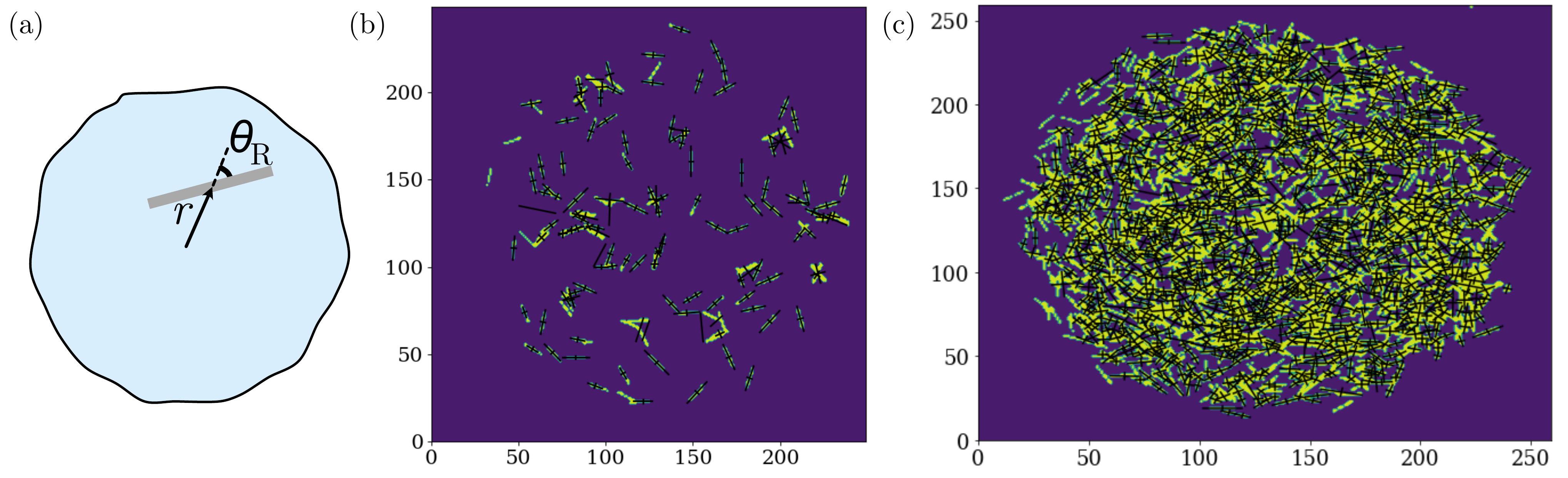}
 \caption{ (a) Schematic of how the orientation $\theta_{\rm R}$ of the fibers is defined (as described in \S3.5 in the main article). (b) A snapshot of the output from the routine used to calculate $\theta_{\rm R}$ for the fiber $A = 12$ and $\phi = 0.1\%$ at $\rm We = 135$, and (c) for the fiber $A = 12$ and $\phi = 5\%$ at $\rm We = 240$. }
 \label{fgr:SM_Fig10}
  \end{center}
\end{figure}

The routine used for identifying the orientation of the fibers relies on a modified version of the fit-ellipse function from the \texttt{opencv} libraries. In addition, we use the \texttt{skimage.segmentation} from \texttt{scikit-image} libraries \cite{van2014scikit}, which enables us to distinguish between overlapping fibers with reasonable accuracy. An example output of this code is shown in figure \ref{fgr:SM_Fig10} (b) and (c). The green rods are the fibers identified from binarization, and the black cross represents the major and minor axis of the fiber fitted with an ellipse.

\smallskip

The segmentation algorithm has its limitations, which result in some uncertainty in the detection of some fibers. However, on average, the routine identifies almost all the fibers visible. While the routine can be verified for dilute suspensions manually, it becomes more complicated to manually count them when the volume fraction increases; for instance, for $\phi \geq 0.5\%$, each droplet contains over 100 fiber. The routine also identifies the center of the droplet automatically and calculates the radial orientation $\theta_{\rm R}$ for all the identified fibers.

\smallskip

Using this approach, we can resolve the orientation of the fibers for volume fraction $\phi \leq 5\%$ with good resolution. For $\phi > 5\%$, however, it is not possible to resolve the individual fibers from the images using segmentation, and the final coating of the fibers essentially resembles a dark puddle after binarization.

\newpage

\section{Isotropy Factor for different $A$}

\begin{figure}[h!]
 \begin{center}
 \includegraphics[width = 0.6\textwidth]{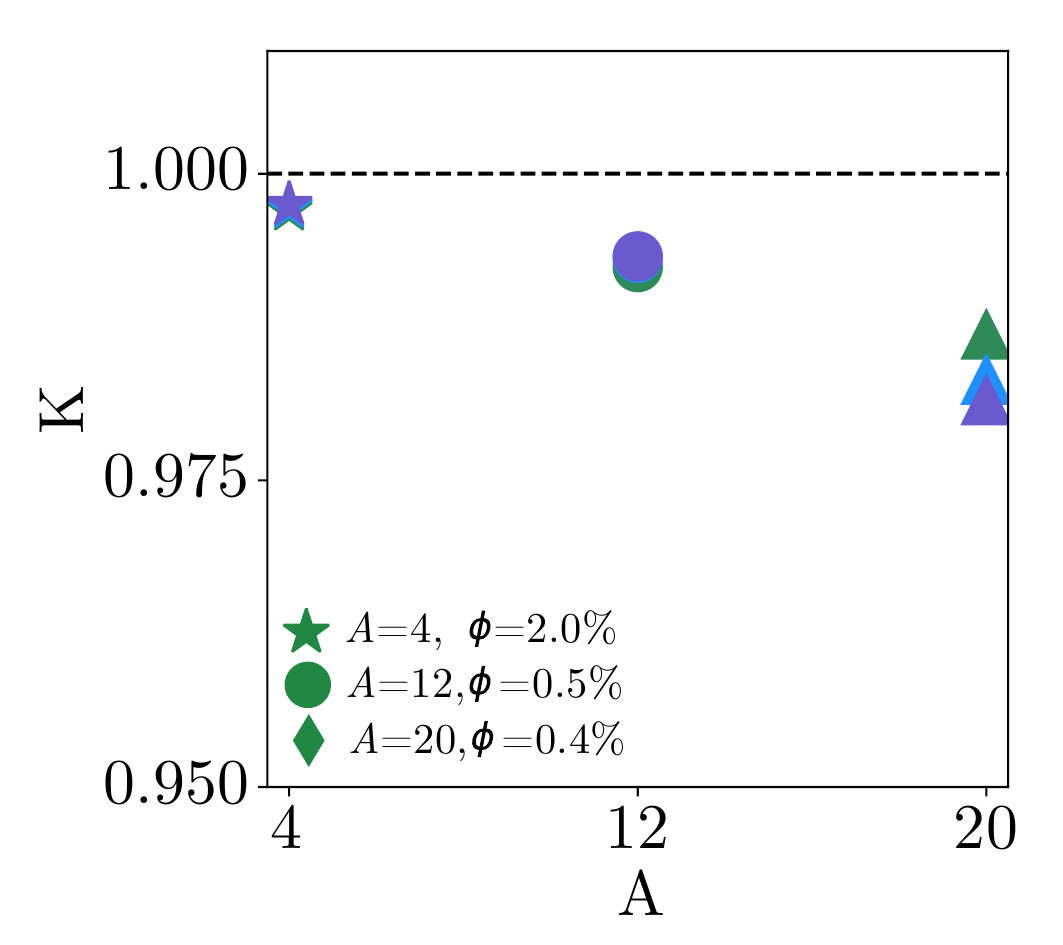}
 \caption{Evolution of the isotropy factor $K$ with the aspect ratio $A$ for fiber suspensions with $\phi = 2\%$, $A = 4$, $\phi = 0.5\%$, $A = 12$ and $\phi = 0.4\%$, $A = 20$.}
 \label{fgr:SM_Fig11}
  \end{center}
\end{figure}

The isotropy factor, defined as $K = 1 - \sigma$, is estimated for droplets of fiber suspension for $A =$ 4, 12, and 20 aspect ratio. The result is reported in figure \ref{fgr:SM_Fig11}. As stated in the main article, $\sigma$ is the standard deviation of the distribution for $\theta_{R}$. From figure \ref{fgr:SM_Fig11}, we see that as $A$ increases, the distribution starts to develop a slight anisotropy. We primarily attribute the rise of anisotropy to a decrease in fiber-fiber interaction rather than viscous dissipation. However, further experiments beyond the scope of the present study will be needed to explore this in detail.

\clearpage

\bibliographystyle{plain}
\bibliography{Biblio}